\documentclass[11pt]{article}
\usepackage{amsfonts}
\usepackage{dsfont}
\usepackage{amssymb}
\usepackage{amsmath}
\usepackage{amsxtra}
\usepackage{graphicx}
\usepackage{psfrag}
\usepackage{mathrsfs}
\usepackage[super]{natbib}
\usepackage{stmaryrd}
\usepackage{subfigure}
\usepackage{tikz}
\usepackage{empheq}
\usepackage[normalem]{ulem}
\usepackage{pdfpages}
\usepackage{color}
\usepackage[superscript]{cite}
\usepackage{longtable}
\usepackage{footmisc}
\usepackage{hyperref}
\usepackage{lineno}

\begin{document}
\def\BGamma{\mbox{\boldmath$\Gamma$}}
\def\BDelta{\mbox{\boldmath$\Delta$}}
\def\BTheta{\mbox{\boldmath$\Theta$}}
\def\BLambda{\mbox{\boldmath$\Lambda$}}
\def\BXi{\mbox{\boldmath$\Xi$}}
\def\BPi{\mbox{\boldmath$\Pi$}}
\def\BSigma{\mbox{\boldmath$\Sigma$}}
\def\BUpsilon{\mbox{\boldmath$\Upsilon$}}
\def\BPhi{\mbox{\boldmath$\Phi$}}
\def\BPsi{\mbox{\boldmath$\Psi$}}
\def\BOmega{\mbox{\boldmath$\Omega$}}
\def\Balpha{\mbox{\boldmath$\alpha$}}
\def\Bbeta{\mbox{\boldmath$\beta$}}
\def\Bgamma{\mbox{\boldmath$\gamma$}}
\def\Bdelta{\mbox{\boldmath$\delta$}}
\def\Bepsilon{\mbox{\boldmath$\epsilon$}}
\def\Bzeta{\mbox{\boldmath$\zeta$}}
\def\Beta{\mbox{\boldmath$\eta$}}
\def\Btheta{\mbox{\boldmath$\theta$}}
\def\Biota{\mbox{\boldmath$\iota$}}
\def\Bkappa{\mbox{\boldmath$\kappa$}}
\def\Blambda{\mbox{\boldmath$\lambda$}}
\def\Bmu{\mbox{\boldmath$\mu$}}
\def\Bnu{\mbox{\boldmath$\nu$}}
\def\Bxi{\mbox{\boldmath$\xi$}}
\def\Bpi{\mbox{\boldmath$\pi$}}
\def\Brho{\mbox{\boldmath$\rho$}}
\def\Bsigma{\mbox{\boldmath$\sigma$}}
\def\Btau{\mbox{\boldmath$\tau$}}
\def\Bupsilon{\mbox{\boldmath$\upsilon$}}
\def\Bphi{\mbox{\boldmath$\phi$}}
\def\Bchi{\mbox{\boldmath$\chi$}}
\def\Bpsi{\mbox{\boldmath$\psi$}}
\def\Bomega{\mbox{\boldmath$\omega$}}
\def\Bvarepsilon{\mbox{\boldmath$\varepsilon$}}
\def\Bvartheta{\mbox{\boldmath$\vartheta$}}
\def\Bvarpi{\mbox{\boldmath$\varpi$}}
\def\Bvarrho{\mbox{\boldmath$\varrho$}}
\def\Bvarsigma{\mbox{\boldmath$\varsigma$}}
\def\Bvarphi{\mbox{\boldmath$\varphi$}}
\def\bone{\mbox{\boldmath$1$}}
\def\bzero{\mbox{\boldmath$0$}}
\def\bnabla{\mbox{\boldmath$\nabla$}}
\def\bvarepsilon{\mbox{\boldmath$\varepsilon$}}
\def\bA{\mbox{\boldmath$ A$}}
\def\bB{\mbox{\boldmath$ B$}}
\def\bC{\mbox{\boldmath$ C$}}
\def\bD{\mbox{\boldmath$ D$}}
\def\bE{\mbox{\boldmath$ E$}}
\def\bF{\mbox{\boldmath$ F$}}
\def\bG{\mbox{\boldmath$ G$}}
\def\bH{\mbox{\boldmath$ H$}}
\def\bI{\mbox{\boldmath$ I$}}
\def\bJ{\mbox{\boldmath$ J$}}
\def\bK{\mbox{\boldmath$ K$}}
\def\bL{\mbox{\boldmath$ L$}}
\def\bM{\mbox{\boldmath$ M$}}
\def\bN{\mbox{\boldmath$ N$}}
\def\bO{\mbox{\boldmath$ O$}}
\def\bP{\mbox{\boldmath$ P$}}
\def\bQ{\mbox{\boldmath$ Q$}}
\def\bR{\mbox{\boldmath$ R$}}
\def\bS{\mbox{\boldmath$ S$}}
\def\bT{\mbox{\boldmath$ T$}}
\def\bU{\mbox{\boldmath$ U$}}
\def\bV{\mbox{\boldmath$ V$}}
\def\bW{\mbox{\boldmath$ W$}}
\def\bX{\mbox{\boldmath$ X$}}
\def\bY{\mbox{\boldmath$ Y$}}
\def\bZ{\mbox{\boldmath$ Z$}}
\def\ba{\mbox{\boldmath$ a$}}
\def\bb{\mbox{\boldmath$ b$}}
\def\bc{\mbox{\boldmath$ c$}}
\def\bd{\mbox{\boldmath$ d$}}
\def\be{\mbox{\boldmath$ e$}}
\def\bff{\mbox{\boldmath$ f$}}
\def\bg{\mbox{\boldmath$ g$}}
\def\bh{\mbox{\boldmath$ h$}}
\def\bi{\mbox{\boldmath$ i$}}
\def\bj{\mbox{\boldmath$ j$}}
\def\bk{\mbox{\boldmath$ k$}}
\def\bl{\mbox{\boldmath$ l$}}
\def\bm{\mbox{\boldmath$ m$}}
\def\bn{\mbox{\boldmath$ n$}}
\def\bo{\mbox{\boldmath$ o$}}
\def\bp{\mbox{\boldmath$ p$}}
\def\bq{\mbox{\boldmath$ q$}}
\def\br{\mbox{\boldmath$ r$}}
\def\bs{\mbox{\boldmath$ s$}}
\def\bt{\mbox{\boldmath$ t$}}
\def\bu{\mbox{\boldmath$ u$}}
\def\bv{\mbox{\boldmath$ v$}}
\def\bw{\mbox{\boldmath$ w$}}
\def\bx{\mbox{\boldmath$ x$}}
\def\by{\mbox{\boldmath$ y$}}
\def\bz{\mbox{\boldmath$ z$}}
\newcommand*\mycirc[1]{%
  \begin{tikzpicture}
    \node[draw,circle,inner sep=1pt] {#1};
  \end{tikzpicture}
}
\newcommand{\upcite}[1]{\textsuperscript{\textsuperscript{\cite{#1}}}}

\makeatletter
\def\@biblabel#1{#1.}
\makeatother
\title{A multi-physics battery model with particle scale resolution of  porosity evolution driven by intercalation strain and electrolyte flow}
\author{Z. Wang\thanks{Department of Mechanical Engineering, University of Michigan}, and K. Garikipati\thanks{Departments of Mechanical Engineering, and Mathematics, University of Michigan, corresponding author, {\tt krishna@umich.edu}}}
\maketitle

\begin{abstract}
We present a coupled continuum formulation for the electrostatic, chemical, thermal, mechanical and fluid physics in battery materials. Our treatment is at the particle scale, at which the active particles held together by carbon-binders, the porous separator, current collectors and the perfusing electrolyte are explicitly modeled. Starting with the description common to the field, in terms of reaction-transport partial differential equations for ions, variants of the classical Poisson equation for electrostatics, and the heat equation, we introduce solid-fluid interaction to the problem. Our main contribution is to model the electrolyte as an incompressible fluid driven by elastic, thermal and lithium intercalation strains in the active material. Our treatment is in the finite strain setting, and uses the Arbitrary Lagrangian-Eulerian (ALE) framework to account for mechanical coupling of the solid and fluid. {\color{black}We present a detailed computational study of the influence of solid-fluid interaction and magnitude of intercalation strain upon porosity evolution, ion distribution and electrostatic potential fields in the cell.}
\end{abstract}
%
%
\section{Introduction}
During battery operation, the intercalation of lithium, thermal and elastic strains drive volume changes in the active material of battery electrodes. As the active material deforms, the porous microstructure of the composite electrode also evolves, and can have a pronounced effect on the effective conductivity, diffusivity and reaction rates at the homogenized, electrode scale. Models at this scale, which is {\color{black} typically} large relative to the pore size of the electrode, assume that the microstructure of the composite electrode (made up of active particles, carbon-binder and fluid electrolyte) does not evolve. A few continuum porous electrode models have been demonstrated, which have a parametric variation of porosity across simulations\cite{White2000JPS,White2010JES}.  However, the coupled physics wherein evolving microstructure drives the variation of effective quantities during battery operation has remained beyond the scope of these studies. Rieger et al.\cite{Rieger2016JES} modeled the expansion of active material particles due to intercalation by assuming the stress to be linearly dependent on lithium concentration. The authors extended this model to the electrode by maintaining a constant volume fraction. Recently, Wang et al. \cite{ZWJES2017} proposed an evolving porosity model at finite strains and studied its effect on battery performance. However, that study depended upon empirical response functions fitted to experimental data for material deformation due to lithium intercalation.

{\color{black}At the particle scale, the deforming active material drives creeping flows of the surrounding electrolyte. In turn, the flow induces a pressure field on the active material, polymeric separator particles and the current collector. Due to the solid-fluid interaction, the porous microstructure of the electrode evolves in a non-uniform manner. It has long been recognized that the microstructure of the electrode can have a pronounced effect on the effective conductivity, diffusivity and reaction rates measured at the homogenized electrode scale. Calculations of transport parameters based on miscrostructure reconstructed using X-ray tomography\cite{Thiele2015, Zielke2014, menetrier1999} have shown that the volume fraction of active material and its grain size, the pore sizes and tortuosity of the pore space have significant effects on these transport parameters. {\color{black}Mistry et al \cite{Mukherjee2018AMI} studied the effects of the conductive binder domain (CBD) in the electrode microstructure by constructing a variety of microstructures with prescribed volume fractions of different solid phases.} Clearly, {\color{black} the numerical study of solid-fluid interaction and the evolution of the microstructure at the particle scale} relies on models that fully resolve the microstructure of the electrodes and separator, including electrolyte flow. Particle scale models\cite{WangJES2007, GoldinEA2012} have been developed to study the effects of different packings of the active material on battery performance. However, only the electrochemical equations were solved over the active material and electrolyte. Malav\'e et al. \cite{MalaveEA2014} proposed a coupled electrochemical-mechanical model to study the mechanical behavior of a single cathode particle. Their model was based on linearized elasticity, thus neglecting the regime of larger intercalation strains (greater than 0.1 \%), with an additive decomposition of the total strain into  elastic  and intercalation components. A similar electrochemical-mechanical model\cite{RobertsJES2014, TrembackiJES2017} has also been applied on the whole battery cell incorporating a conductive binder. {\color{black} Mendoza et al.\cite{MendozaEA2016} included the stress dependent chemical potential in their model for lithium transport in the particle phase}. A more sophisticated mechanical model\cite{RahaniJES2013} was explored to study the plastic deformation of only the binder via a non-linear constitutive relationship for the stress-strain response. {\color{black} Crack formation inside active particles was modeled using a single particle model framework\cite{KotakJES2018} to study the mechanical degradation of lithium ion batteries.}
In related  work, reconstruction of the microstructure from X-ray tomography data has made simulations on realistic electrodes possible.\cite{Fang2006, GoldinEA2012, RobertsJES2014}. We note the work of Wang and Sastry \cite{WangJES2007} who studied the effective diffusivity and capacity for various idealized, close-packed, as well as experimentally imaged three-dimensional microstructures. Their work, however,  neglected any mechanical influences. Fang et. al\cite{Fang2006} proposed a coupling approach based on the mortar finite element method that allows non-matching meshes at the electrode-electrolyte interfaces while solving the electro-chemical equations. Mechanical influences were included in the particle-scale model of Rahani and Shenoy \cite{RahaniJES2013}, who incorporated the plastic deformation of the binder to study stress evolution during battery cycling. However, it has not been common to see the effects of mechanics  coupled back into the electro-chemical behavior of electrodes at the particle scale. A recent step in this direction was taken by Trembacki and co-workers \cite{TrembackiJES2017}, who carried out finite element computations on experimentally determined microstructures by including the effect of mechanics on conductivity.}

Previous modelling work, including those cited above, did not explicitly consider the porous separator material; the separator was represented as a {\color{black} a homogenized medium with effective properties}. The fluid motion of the electrolyte was neglected. While the mathematical descriptions of mass and momentum conservation in the electrolyte have been presented \cite{WangJES1998,TorabiJES2011}, {\color{black} direct numerical simulation of the coupling of these effects} has been limited. As mentioned previously, the electrolyte undergoes creeping flow past the active particles and polymeric separator's microstructure \cite{GuJES2000, XuPECS2015}. Modelling the electrode and separator as porous media, the momentum equation reduces to Darcy's law. At the particle scale, Qiu et al.\cite{QiuEA2012} applied the lattice Boltzmann method to avoid solving the Navier-Stokes equations in complex geometries for modeling flow of the liquid electrolyte. However, the mechanical response of the solid component was not included in these models, and solid-fluid interactions therefore could not be studied. 

To be best of our knowledge, there is no complete computational treatment for lithium-ion batteries that incorporates electrolytic flow coupled with the electro-chemo-thermo-mechanical phenomena at the particle scale. In this communication, we aim to present such a model by combining the following features: (a) explicitly modeling the electro-chemo-thermo-mechanical equations over the current collectors, active material, conductive carbon-binder, porous separator and electrolyte making up the cell; (b) electrolytic flow  modelled as an incompressible fluid; (c) the lithium concentration and temperature fields that drive inelastic components of strain governed by the equations of nonlinear elasticity; (d) coupled solid-fluid interactions. We first present an extension of the framework of Newman and Thomas-Alyea \cite{NewmanBook} to account for the flow velocity of the  electrolyte in the species transport and thermal equations {\color{black}(cf. Section \ref{sec:strong form})}. We then discuss the interface conditions for electro-chemistry and solid-fluid mechanics at every solid-fluid internal boundary {\color{black}(cf. Section \ref{sec:BC})}. 
Numerical and computational issues are discussed {\color{black}(cf. Section \ref{sec:Numerical})}, before
studying the effects of different far-field boundary conditions, and of evolving porosity on the performance of the cell
 {\color{black}(cf. Section \ref{sec:Results})}. Concluding remarks appear under Discussion and conclusions {\color{black}(cf. Section \ref{sec:Conclusions})}.

{\color{black}The open source code used in this work is available in a user-friendly form at \url{https://github.com/mechanoChem/mechanoChemFEM/tree/example}.}

\section{The multi-physics particle scale model}
\label{sec:strong form} 
\begin{figure}[hbtp]
\centering
\includegraphics[scale=0.3]{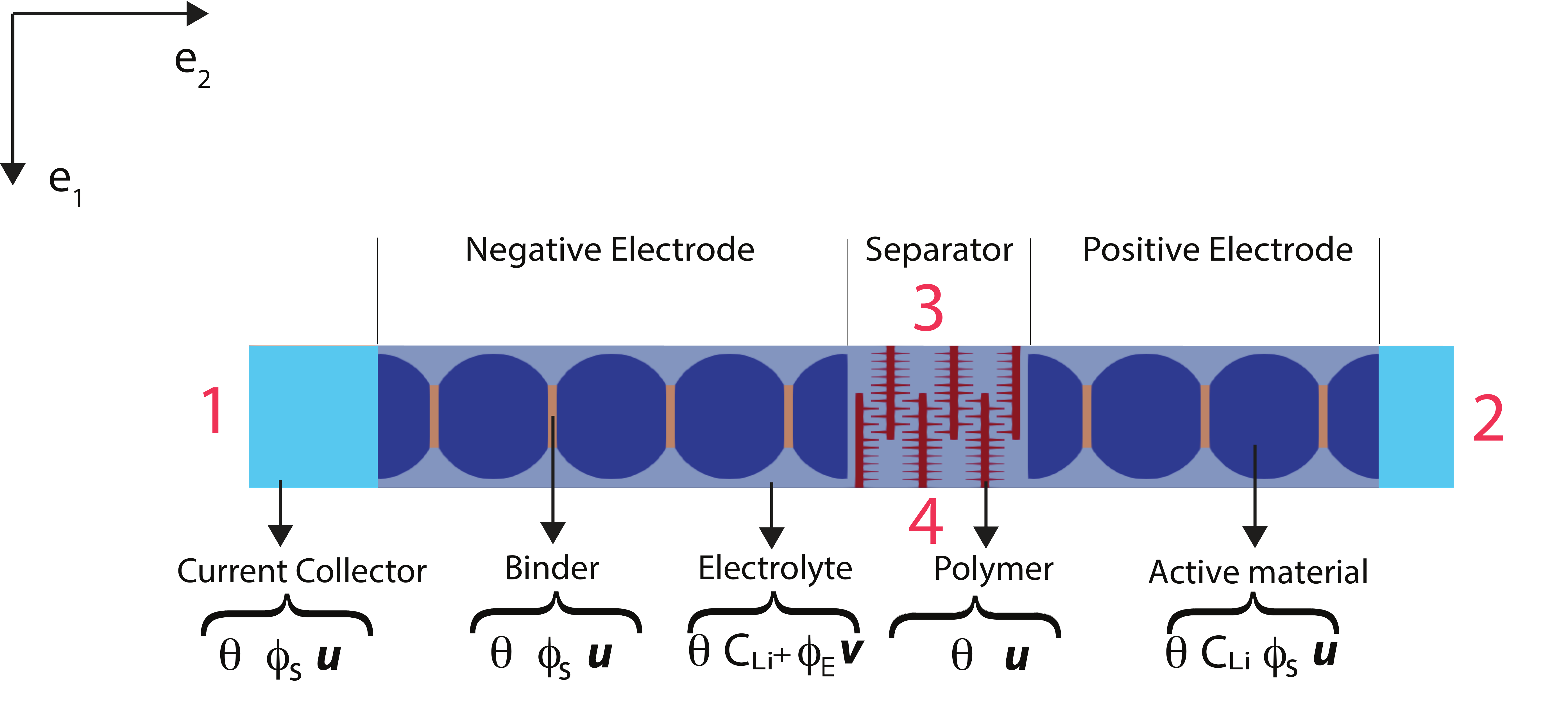}
\caption{{\color{black}A schematic showing each component of a battery cell, as considered in the model, with active materials represented as the union of intersecting spherical particles aligned along the electrode thickness ($\be_2$ direction) and connected by carbon-binder additive. Particles of both electrodes share the same diameter. The entire battery cell is a rolled or folded structure of a layer that is very thin in the $\be_2$ direction compared with the two in-plane directions $\be_1$ and $\be_3$. Here we show a section perpendicular to the $\be_3$ direction with each boundary surface labeled.} }
\label{fig:schematicParticleModel}
\end{figure}

A battery cell consists of porous, positive and negative electrodes, a separator and current collectors (see Figure \ref{fig:schematicParticleModel}). The electrode consists of active-material particles and carbon-binders, and the pore space is filled with the electrolyte. The two electrodes are isolated by a separator which is usually made of polyolefin for Li-ion batteries. The porous separator is perfused with electrolyte, allowing ionic transport between the electrodes. Metallic current collectors are located at either end of the battery.

We lay down the governing equations for primary variables in three dimensions. The electro-chemical treatment follows the pioneering work of Newman and Thomas-Alyea\cite{NewmanBook}. It is then coupled with the thermal field governed by the heat equation, with heat production from charge transport and reactions. The novel aspect of our framework is the incorporation of mechanics at finite strain, driven by lithiation- and temperature-induced swelling, and explicit modeling of electrolytic flow as an incompressible fluid. As is the convention in continuum mechanics, the solids (current collector, active-material particle and polymeric separator particles)
are described in a Lagrangian setting, {\color{black}while the transport equations for ions and heat}, as well as the fluid (electrolyte) flow are described in an Eulerian setting. {\color{black} However, we do not reproduce the well-known derivation of the equations. Because advection velocities are low, we also leave those terms out of consideration.} Electrolytic flow around the deforming solid components alters the fluid domain. The fluid mesh therefore must be remapped as the computation proceeds. The Arbitrary Lagrangian-Eulerian (ALE) framework is adopted here for this purpose.

\subsection{Electro-chemo-thermal equations}
The electro-chemical equations presented here differ from the work of Newman and Thomas-Alyea\cite{NewmanBook} in that all variables are defined in the deformed configuration. As explained above, this requires the ALE framework to solve these equations over the fluid (See Section \ref{sec:ALE}). 

\noindent In the active material we have conservation of lithium:

\begin{align}
\frac{\partial C_\text{Li}}{\partial t}+\nabla\cdot\bj=0
\label{eq:ConsCli}
\end{align}
\begin{align}
\bj=-D\nabla C_\text{Li}
\end{align}

\noindent where $D$ is the diffusivity. {\color{black}As is well-known, mass fluxes are driven by a generalized chemical potential gradient, which includes contributions from concentration, stress and electric field gradients \cite{degrootmazur1984}. Other authors have considered these effects in battery materials \cite{MendozaEA2016, BowerJMPS2011}, as we have for more general treatments of mechano-chemical coupling \cite{Rudrarajuetal2016}. Here, we neglect other terms  and reduce the form of the flux to directly express it in terms of the concentration gradient.}
We model the electrolyte as a binary liquid, in which the conservation of of Li$^+$  cations is also written as:
\begin{align}
\frac{\partial C_{\text{Li}^+}}{\partial t}+\nabla\cdot\bj_+=0
\label{eq:ConsCli+}
\end{align}
In general, the flux of each species can depend on the concentration gradients of the other species. This is a special case of the Onsager reciprocity relations \cite{degrootmazur1984} with the dependence on chemical potential gradients reduced to concentration gradient dependence. Following Newman and Thomas-Alyea\cite{NewmanBook}, we have 
\begin{align}
\bj_+=-D_{+}\nabla C_{\text{Li}^+} +\frac{t_+}{F}\bi_\text{E}+C_{\text{Li}^+}\bv
\label{eq:Cli+flux}
\end{align}
for a binary solution. Here, $D_{+}$ is the diffusivity, $t_+$ is the transference number of the cation, $\bv$ is the electrolyte velocity, $F$ is the Faraday constant, and $\bi_\text{E}$ is the total current in the electrolyte phase which will be derived in what follows.

The total current in the current collector, conductive carbon-binder and active material is governed by Ohm's law:
\begin{align}
\bi_\text{S}=-\sigma_\text{S} \nabla\phi_\text{S}
\label{eq:currentDensitySolid}
\end{align}
where $\phi_\text{S}$ is the electric potential, the subscript stands for ``solid'', and the conductivity $\sigma_\text{S}$ differs between the current collector and active material. In the electrolyte we have
\begin{align}
\bi_\text{E}=-\sigma_\text{E}\nabla\phi_\text{E}-\gamma_\text{D}\nabla\ln C_{\text{Li}^+}
\label{eq:currentDensityElectrolyte}
\end{align}
where  $\sigma_\text{E}$ is the electrolyte's conductivity, and $\gamma_\text{D}$ is the diffusion conductivity evaluated as in Goldin et al. \cite{GoldinEA2012}:
\begin{align}
\gamma_\text{D}=\frac{2R\theta \sigma_\text{E}}{F}\left(1-t_+\right)\left(1+\frac{d \ln f}{d\ln  C_{\text{Li}^+}}\right),
\label{eq:diffusionConductivityFull}
\end{align}
{\color{black}where $R$ is the universal gas constant, $\theta$ is the temperature}, and $f$ is the mean molar activity coefficient of the electrolyte, which is assumed to be constant. Thus, the relation simplifies to
\begin{align}
\gamma_\text{D}=\frac{2 R\theta \sigma_\text{E}}{F}\left(1-t_+\right).
\label{eq:diffusionConductivity}
\end{align}

During charging, the dissociation $\mathrm{Li} \rightarrow \mathrm{Li}^+ + \mathrm{e}^-$ at the cathode-electrolyte interface gives rise to a layer of negative charge on the active material and a layer of positive charge in the electrolyte. The reverse association reaction $ \mathrm{Li}^+ + \mathrm{e}^-\rightarrow \mathrm{Li}$ creates a layer of negative charge in the electrolyte and a layer of positive charge on the active material. However, the thickness of this  layer, the Debye length $\sim 1$ nm, {\color{black} \cite{Dickinson2011}} is often neglected at larger length scales, in favor of the electroneutrality approximation. More details can be found in Newman and Thomas-Alyea \cite{NewmanBook}, and a historical perspective in Dickinson et al.\cite{Dickinson2011}.
With the electroneutrality approximation, conservation of charge leads to
\begin{align}
\nabla\cdot\sum z_i \bj_i=0.
\end{align}
{\color{black}where $z_i$ is the charge number.} Note that, for a binary electrolyte, the summation accounts for the flux of cations as shown in Equation (\ref{eq:Cli+flux}), and the flux of anions, which are not explicitly modeled here due to the electroneutrality approximation.
The total current is due to the motion of charged particles, giving:
\begin{align}
\bi=F\sum z_i\bj_i.
\label{eq:current}
\end{align}
That is, 
\begin{align}
 \nabla\cdot \bi=0
 \label{eq:consCurrent}
 \end{align}
 under the electroneutrality approximation.
From Equations (\ref{eq:currentDensitySolid}) and  (\ref{eq:currentDensityElectrolyte}) , we have
\begin{align}
\nabla\cdot (- \sigma_\text{s}\nabla\phi_\text{S})=0
\label{eq:reducedPhis}
\end{align}
in the active material, and
\begin{align}
\nabla\cdot\left(-\sigma_\text{E}\nabla\phi_\text{E}-\gamma_\text{D}\nabla\ln C_{\text{Li}^+}\right)=0
\label{eq:reducedPhie}
\end{align}
in the electrolyte. These two equations properly describe the electric potentials with the electroneutrality approximation, such that the double layer effect is neglected.

\subsection{The standard thermal equations}
{\color{black}We solve for the temperature in order to account for its effect on the coefficients of the electro-chemical equations: These include the lithium ion diffusivity (Equation \ref{eq:D_plus}), conductivity (Equation \ref{eq:diffusionConductivityFull}) and reaction parameters (Equation \ref{eq:BVeuqation})}.
Heat generation and transport are governed by the heat equation, which is derived from the first law of thermodynamics. {\color{black} While we do model flow of the electrolyte in Section \ref{sec:flow}, its velocity is low. We estimate that the Peclet number, {\color{black}$\text{Pe}=\frac{v\rho C_p L}{\lambda} \sim 1.0\times 10^{-8}$} from our computations. Therefore we neglect heat flux by advection.} For the temperature $\theta$, we have the standard form of the heat equation\cite{Bernardi1985}:
\begin{align}
\rho C_p\frac{\partial\theta}{\partial t}+\nabla \cdot\bq=0
\label{eq:heat}
\end{align}
where $\rho$ is  the mass density of the electrode,  $C_p$ is the specific heat and $\bq$ is the heat flux. The heat flux can be expressed as
\begin{align}
\bq=\phi F\sum_i z_i\bj_i-\lambda\nabla\theta+\bq^\text{D}
\label{eq:heatflux}
\end{align}
where $\lambda$ is the thermal conductivity. The first term on right is associated with Joule heating and the last term $\bq^\text{D}$ is the Dufour effect. {\color{black}For binary liquid mixtures, such as the electrolyte considered here, the Dufour effect is usually negligible, and is ignored in this work \cite{NewmanBook}}. Again, with the electroneutrality approximation we have $\nabla\cdot\sum z_i \bj_i=0$, and substituting Equation (\ref{eq:heatflux}) into the heat equation (\ref{eq:heat}), we have
\begin{align}
\rho C_p\frac{\text{d}\theta}{\text{d}t}-\lambda \nabla^2\theta+\nabla\phi_\text{S}\cdot\bi_\text{S}=0
\label{eq:heatElectrode}
\end{align}
in the electrode, and 
\begin{align}
\rho C_p\frac{\text{d}\theta}{\text{d}t}-\lambda \nabla^2\theta+\nabla\phi_\text{E}\cdot\bi_\text{E}=0
\label{eq:heatElectrolyte}
\end{align}
in the electrolyte. 
%
%
\subsection{Finite strain mechanics}
\label{sec:mechanics}
Lithium intercalation and de-intercalation induce expansion and contraction, respectively, of the active material. Additionally, the active material, carbon-binder, porous separator and current collector undergo thermal expansion. 
The kinematics of finite strain leads to the following decomposition in the active material:
\begin{align}
\bF=\bF^\text{e}\bF^\text{c}\bF^{\theta}.
\label{eq:cellDeformGradientParticle}
\end{align}
For other solid sub-domains (carbon-binder and polymeric separator particles) we have
\begin{align}
\bF=\bF^\text{e}\bF^{\theta}.
\label{eq:cellDeformGradient}
\end{align}
Here, $\bF = \bone + \partial\bu/\partial\bX$, is the total deformation gradient tensor in each of the relevant solid regions. Its multiplicative components $\bF^\text{e}$, $\bF^\text{c}$ and $\bF^{\theta}$, are, respectively, the elastic, chemical (induced by lithium intercalation) and thermal components. In the absence of a body force the strong form of the mechanics problem in the current configuration is 
\begin{align}
\nabla\cdot\bT = \bzero,
\label{eq:strongformElasticity}
\end{align}
\begin{align}
\text{for}\quad \bT= \frac{1}{\det{\bF^\text{e}}}\frac{\partial W}{\partial \bF^\text{e}}\bF^{\text{e}^\text{T}},
\end{align}
where $\bT$  is the Cauchy stress tensor and $W$ is the strain energy density function {\color{black}which is chosen to be the Saint Venant–-Kirchhoff model:
\begin{align}
W(\bE)&=\frac{\lambda}{2}[\text{tr}(\bE)]^2+\mu \text{tr}(\bE^2)\\
\text{where} \quad \bE &= \bF^{\text{e}^\text{T}}\bF^\text{e}
\end{align}
and $\lambda$ and $\mu$ are  the Lam\'{e} constants.}
The chemical and thermal expansion components of the multiplicative decomposition of $\bF$ are modelled as isotropic, e.g.:
\begin{align}
F^{\text{c}}_{iJ}=(1+\beta^{\text{c}})^{1/3}\delta_{iJ} \label{eq:deformFc}\\
F^{\theta}_{iJ}=(1+\beta^{\theta})^{1/3}\delta_{iJ} \label{eq:deformFT}
\end{align}
where $\delta_{iJ}$ is the Kronecker delta. The lithiation swelling response function, {\color{black}$\beta^{\text{c}}$}, is parameterized by the lithium concentration, $C_\text{Li}$, and can be fit to data \cite{ZWJES2017} as shown in Equation (\ref{eq:beta_s}). We write
\begin{align}
\beta^{\theta}(\theta)&=\Omega_{\theta} (\theta-\theta_0)
\end{align}
for the thermal expansion functions.

{\color{black}Since the dimension in the $\be_3$ direction (see Figure \ref{fig:schematicParticleModel}) is very large compared to the others, we adopt the plane strain assumption in the majority of our computations. Strictly speaking, this also implies that the particles are cylindrical, with their axes in the $\be_3$-direction, which is a simplification that ignores the three-dimensionality of the microstructure. We also do demonstrate the fully three-dimensional case in the Appendix.} 

\subsection{Incompressible fluid model}
\label{sec:flow}
We model the electrolyte as an incompressible, creeping flow.  \footnote{\color{black} We assume that higher-order charge effects are absent, such as the density of electrolyte depending on the lithium ion concentration.} In this regime, with inertia and body forces being neglected, the Stokes equations are: 
  \begin{align}
-2\eta\nabla\cdot\varepsilon(\bv)+\nabla p=0
\label{eq:stoke}
 \end{align}
 \begin{align}
\nabla\cdot\bv=0
\label{eq:continueEq}
 \end{align}
where $\Bvarepsilon(\bv)=\frac{1}{2}(\nabla\bv+\nabla\bv^\text{T})$ is the strain rate tensor, and $\eta$ is the dynamic viscosity.

\subsection{The Arbitrary Lagrangian-Eulerian framework}
\label{sec:ALE}
In the solid regions, i.e., the current collector, active-material, carbon-binder and polymeric separator particles, the Lagrangian description attaches the finite element nodes to material points. In the fluid (electrolyte) phase, whose description is Eulerian, the mesh must be mapped to evolve with the  fluid sub-domain as it flows around the deforming solid components. If the mesh displacement is $\bu_\text{m}$, the spatial gradient operator acting on a variable $\tau$ in the Eulerian setting is
\begin{align}
    \nabla \tau=\frac{\partial\tau}{\partial\bX}\bF_\text{m}^{-1}
    \label{eq:ALEspaceDerivetive}
\end{align}
where $\bF_\text{m}= \bone +\partial \bu_\text{m}/\partial \bX$, is the  deformation gradient tensor of the mesh. In the solid phase, $\bu_\text{m}$ coincides with the displacement of material points, $\bu$, and $\bF_\text{m}$ is replaced by the deformation gradient tensor of the solid, $\bF$. In the fluid phase the description of fluid mesh deformation could be arbitrary, but {\color{black} extreme distortions of the mesh may cause numerical difficulties, and the treatment of fluid mesh motion therefore proves to be critical in solid-fluid interaction problems}. The mesh displacement in the fluid phase can be solved by the Poisson equation, arbitrary elasticity equations or bi-harmonic equations. For small mesh deformation, we can simply solve $\bu_\text{m}$ in the fluid phase by solving a linear elasticity problem with arbitrary parameters.
\begin{align}
\nabla\cdot\bT_\text{m} = \bzero
\label{eq:u_mesh}
\end{align}
\begin{align}
\bT_\text{m}= \bC:\Bvarepsilon_\text{m}
\end{align}
where $\bT_\text{m}$ is the ``virtual'' Cauchy stress tensor, $\bC$ is the arbitrary fourth-order stiffness tensor and $\Bvarepsilon_\text{m}(\bu)=\frac{1}{2}\left(\nabla\bu_m+(\nabla\bu_m)^T\right)$ is the infinitesimal strain tensor.

More simply, the mesh displacement field can be solved by the Poisson equation:
\begin{align}
    \nabla^2 \bu_\text{m}=\bzero
    \label{eq:ALEPossion}
\end{align}
However large lithium intercalation strain, $\beta^c$, may lead to a dramatic volume change of the active material. The large deformation of active material may cause extreme distortions of the mesh in the fluid. For this reason, we apply  adaptive mesh rezoning schemes\cite{Masud2007}. The idea is to introduce a constraint condition over two adjacent element nodes $i$ and $j$ such that the relative difference of net displacements is {\color{black} not greater} than the element length $h^e$:
\begin{align}
|\bu_m^i-\bu_m^j|\le\alpha h^e,
\end{align}
which is equivalent to
\begin{align}
    |\nabla\bu_m|\le\alpha
\end{align}
where {\color{black}$\alpha\in[0,1]$} is the tolerance for element distortion. This constraint is applied element-wise and can be easily incorporated into Equation (\ref{eq:ALEPossion}) as a penalty term:
\begin{align}
    \nabla^2(1+\tau_\text{m}\bu_\text{m})=\bzero
    \label{eq:ALEPossion_constrain}
\end{align}
where $\tau_\text{m}$ is the weight function that imposes spatially varying stiffening effects for the mesh. The key idea for choosing $\tau_\text{m}$ is to enforce this value to be high for smaller elements and small for large elements. {\color{black} Consequently, smaller elements are stiffer than larger elements in the mesh. The mesh distorts
non-uniformly, with distortion being concentrated
to the larger elements.}  In this work we choose
\begin{align}
   \tau_m=\left(\frac{\det \bF_\text{m}^0}{\det \bF_\text{m}} \right)^\delta
\end{align}
where $\bF_\text{m}^0$ is the initial deformation gradient tensor of the mesh and $\delta$ is a constant value. {\color{black}Rezoning alone is not likely to be sufficient to enable numerical solutions in cases of extreme intercalation strains, such as the 300\% observed in silicon electrodes. In such cases, the expanding particles drive electrolyte flows that deform the polymeric separator, which can distort the fluid mesh, while itself coming into contact with the extremely strained active particles. See Figure \ref{fig:V_zoom} for an example of such distortion, albeit with a much lower intercalation strain of $\sim 5$\%. Other techniques including remeshing and the invocation of contact mechanics will be necessary to further drive the solution in such cases.}

The time derivatives within the ALE framework also need to be rewritten as
\begin{align}
    \partial_t=\partial_t^\ast-(\bv_\text{m}\cdot \nabla)
\end{align}
where $\partial_t^\ast$ is the ALE time derivative and $\bv_\text{m}$ is the mesh velocity. {\color{black}Note that the term $\bv_\text{m}\cdot \nabla$ represents mesh motion by advection.} For the solid phase $\bv_\text{m}$ coincides with the material deformation rate which is negligible in this problem. Consequently, $\bv_\text{m}$ in the fluid phase, which is driven by the deformation rate in the solid phase is also small, and we use the approximation
\begin{align}
    \partial_t =\partial_t^\ast
    \label{eq:ALEtimeDerivetive}
\end{align}
Thus, to write formulations in the ALE framework, we only need to replace the spatial derivatives and time derivatives by formulas (\ref{eq:ALEspaceDerivetive}) and (\ref{eq:ALEtimeDerivetive}). {\color{black} To the best of our knowledge, the ALE framework has not been previously applied to solid-fluid interaction in particle-scale battery models.}

\section{Boundary and interface condition}
\label{sec:BC}
\begin{figure}[hbtp]
\centering
\includegraphics[scale=0.3]{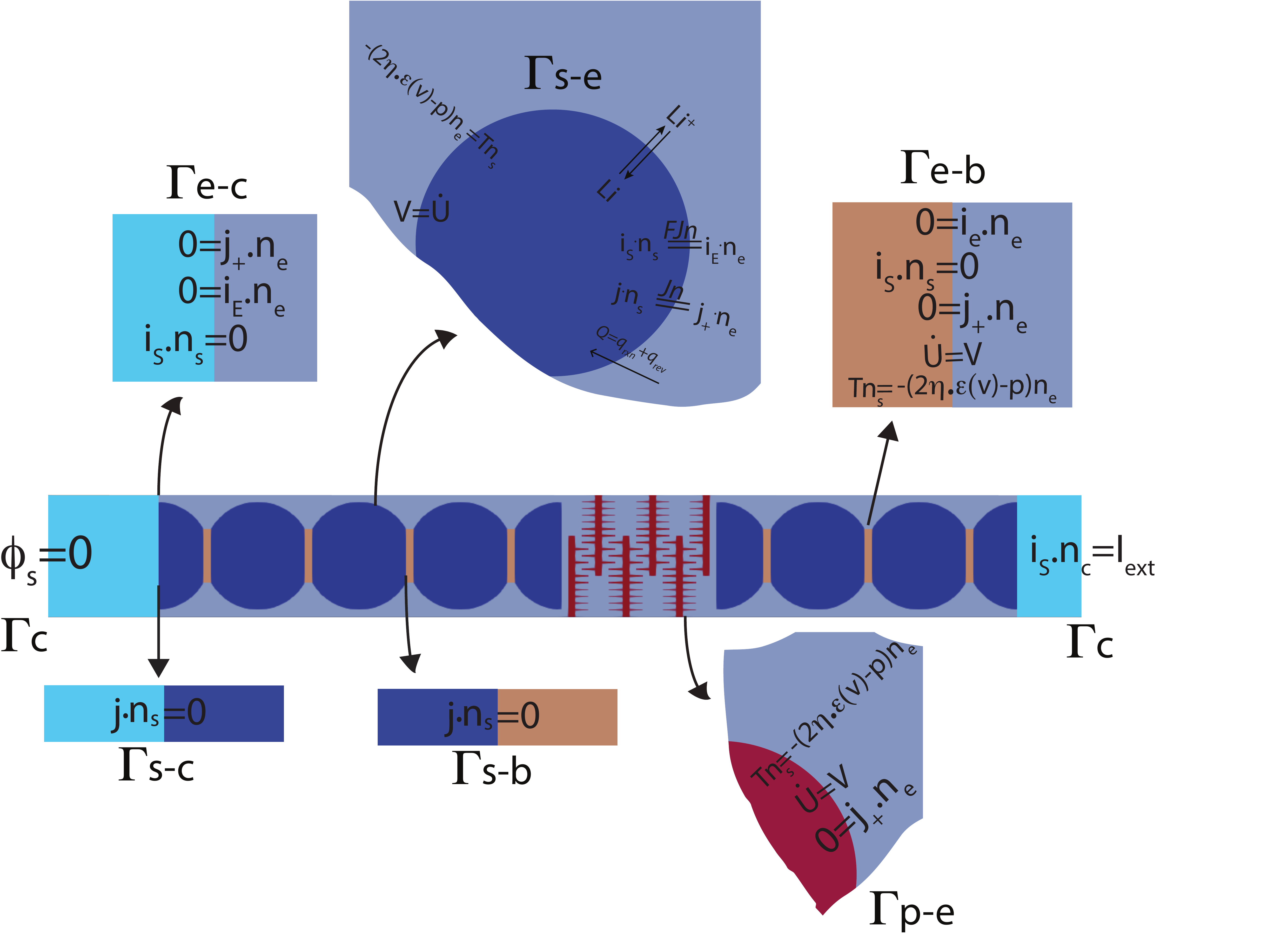}
\caption{\color{black}A schematic showing boundary and interface conditions.}
\label{fig:BC}
\end{figure}

The multi-physics character of this problem extends to the restriction of certain physics--and therefore the corresponding partial differential equations--to specific sub-domains. The conventional application of boundary conditions translates to interface conditions where the sub-domains meet.  We first discuss decoupled interface conditions and domain boundary conditions before the more complicated, coupled interface conditions. The boundary and interface conditions are also shown in Figure \ref{fig:BC}.

\subsection{Decoupled interface conditions, and domain boundary conditions}
Since lithium and lithium ions are absent from the current collector, carbon-binder and polymeric separator particles, we have 
 \begin{align}
 \bj\cdot\bn_\text{s}=0 \quad \text{on } \Gamma_\text{s-c}\\
  {\color{black}\bj\cdot\bn_\text{s}=0 \quad \text{on } \Gamma_\text{s-b}}\\
  \bj_+\cdot\bn_\text{e}=0 \quad \text{on } \Gamma_\text{e-c}\\
    \bj_+\cdot\bn_\text{e}=0 \quad \text{on } \Gamma_\text{e-b}\\
        \bj_+\cdot\bn_\text{e}=0 \quad \text{on } \Gamma_\text{e-p}
 \end{align} 
where $\Gamma_\text{s-c}$ is the interface between the active material and current collector, {\color{black}$\Gamma_\text{s-b}$ is the interface between the active material and carbon-binder}, $\Gamma_\text{e-c}$ is the interface between the electrolyte and current collector, {\color{black} $\Gamma_\text{e-b}$ is the interface between the electrolyte and carbon-binder}, and $\Gamma_\text{e-p}$ is the interface between the electrolyte and polymeric separator particle. Here, $\bn_\text{s}$, $\bn_\text{e}$ are the corresponding outward unit normal vectors from the active material and electrolyte, respectively. Also, since ions do not enter the carbon-binder and current collector, the current in the electrolyte vanishes at $\Gamma_\text{e-b}$ and $\Gamma_\text{e-c}$, leading to 
  \begin{align}
  \bi_\text{E}\cdot\bn_\text{e}=0 \quad \text{on }\Gamma_\text{e-b}\\
 \bi_\text{E}\cdot\bn_\text{e}=0 \quad \text{on } \Gamma_\text{e-c}
 \end{align}
The external current  applied at the domain boundary on the current collector is
  \begin{align}
 \bi_\text{S}\cdot\bn_\text{c}=\bi_\text{ext} \quad \text{on } \Gamma_\text{c}
 \end{align}
 where $\Gamma_\text{c}$ is the outer boundary of the current collector and $\bn_\text{c}$ is the outward unit normal vector.
 Boundary conditions for the temperature, $\theta$, are in the form of conductive heat transfer to the ambient air, written as
 \begin{align}
 -\lambda \nabla \theta\cdot \bn=h(\theta-\theta_\text{air}) \quad \text{on } \Gamma_\text{outer}
 \end{align}
where $\Gamma_\text{outer}$ is the outer boundary surface. {\color{black}Since, in practice, the battery cells are stacked atop each other, there is no route for heat transfer to the ambient air, which we approximate by applying an adiabatic boundary condition $h=0$ on $\Gamma_\text{outer}$}.
 The above conditions are decoupled, but the primal variables can also be coupled at interfaces as discussed below. 
 
\subsection{Chemical reaction at the interface between active material and electrolyte}
During discharging or charging, the following chemical reactions occur at the interface between the active material and electrolyte:
\begin{align}
\text{Li}\to \text{Li}^++q^-, \quad \text{-ve electrode during discharge/+ve electrode during charge,}\nonumber\\
 \text{Li}^++q^- \to\text{Li},\quad \text{+ve electrode during discharge/-ve electrode during charge}.\nonumber
\end{align}
The reaction rate is modelled by the
the Butler-Volmer equation
\begin{align}
j_n&=j_0\left(\text{exp}\left(\frac{\alpha_aF}{R\theta}(\phi_\text{S}-\phi_\text{E}-U)\right)-\text{exp}\left(-\frac{\alpha_aF}{R\theta}(\phi_\text{S}-\phi_\text{E}-U)\right)\right) \label{eq:BVeuqation}\\
j_0&=k_0(C_{\text{Li}^+})^{\alpha_a} \frac{(C_\text{Li}^{\text{max}}-C_\text{Li})^{\alpha_a}}{(C_1^{\text{max}})^{\alpha_a}}  \frac{(C_\text{Li})^{\alpha_c}}{{(C_1^\text{max}})^{\alpha_c}}
\label{eq:BVeuqationj0}
\end{align}
where $j_0$ is the exchange current density, $\alpha_a, \alpha_c$ are transfer coefficients and $k_0$ is a kinetic rate constant. The open circuit potential can be written as a fit to the state of charge (SOC)\cite{ZWJES2017}, leading to an interface condition for lithium and lithium ions:
 \begin{align}
\bj\cdot \bn_\text{s}&=j_n \quad \text{on } \Gamma_\text{s-e},\\
\bj_+\cdot \bn_\text{e}&=-j_n\ \quad \text{on } \Gamma_\text{s-e},
 \end{align}
where $j_n$ is the reaction rate of lithium dissociation, which is positive at the negative electrode and negative at the positive electrode during discharging. The active material-electrolyte interface is $\Gamma_\text{s-e}$. The reaction rate is related to the current by $\bi\cdot\bn=Fj_n$, giving:
 \begin{align}
 \bi_\text{S}\cdot\bn_\text{s}&=Fj_n  \quad \text{on } \Gamma_\text{s-e}\\
 \bi_\text{E}\cdot\bn_\text{e}&=-Fj_n \ \quad \text{on } \Gamma_\text{s-e}
 \end{align}
{\color{black} While there is some evidence that ionic conductivity can occur in nanopores of the carbon-binder, and hence, solid state reactions also could occur at these active particle-binder interfaces \cite{Thiele2015}, we do not consider this phenomenon.} 
\noindent At the interface, the chemical reactions also generate heat:
 \begin{align}
Q_{\Gamma_\text{s-e}}=q_{\text{rxn}}+q_{\text{rev}}
 \end{align}
 where
\begin{subequations}
\begin{alignat}{2}
q_{\text{rxn}}&=Fj_n(\phi_\text{S}-\phi_\text{E}-U),  &&\;\text{irreversible entropic heat,} \label{pa:irreversibleentropicheat}\\
q_{\text{rev}}&=Fj_n\theta\frac{\partial U}{\partial \theta}, &&\;\text{reversible entropic heat.} \label{pa:reversibleelectrochemicalreactionheat}
\end{alignat}
\end{subequations}
At the interface, some fraction of $Q_{\Gamma_\text{s-e}}$ enters each of the adjoining sub-domains. {\color{black} Because of the generally higher thermal conductivity of the electrodes over electrolytes \cite{Werner2017}, most of the heat produced at $\Gamma_{\text{s}-\text{e}}$ flows into the active material. For the sake of simplicity, we assume that $Q_{\Gamma_\text{s-e}}$ flows entirely into the active material:}
\begin{align}
-\bq\cdot\bn_\text{s}=q_{\text{rxn}}+q_{\text{rev}}  \quad \text{on } \Gamma_\text{s-e}
\end{align}

\subsection{Solid-fluid interaction}
The electrolytic fluid is bounded within the solid materials: active material, carbon-binder, polymeric separator particles, and current collector. No-slip conditions give  
\begin{align}
\bv=\dot \bu \quad \text{on } \Gamma_\text{s-e}, \Gamma_\text{e-c}, \Gamma_\text{e-p}
\label{eq:interfaceV-u}
\end{align}
Traction continuity on all solid-fluid interfaces gives:
\begin{subequations}
\begin{align}
\bT\bn_\text{s}&=-(2\eta\cdot\varepsilon(\bv)-p\mathds{1})\bn_\text{e} \quad \text{on } \Gamma_\text{s-e}
\label{eq:interfaceU-v-s}\\
\bT\bn_\text{c}&=-(2\eta\cdot\varepsilon(\bv)-p\mathds{1})\bn_\text{e} \quad \text{on } \Gamma_\text{e-c}
\label{eq:interfaceU-v-c}\\
\bT\bn_\text{p}&=-(2\eta\cdot\varepsilon(\bv)-p\mathds{1})\bn_\text{e} \quad \text{on } \Gamma_\text{e-p}
\label{eq:interfaceU-v-p}\\
\end{align}
\end{subequations}

Lithium intercalation and de-intercalation induces expansion and contraction of the active material, which drives the electrolytic flow. Conversely, the flowing electrolyte drives solid material deformation at all solid-fluid interfaces. Compatibility at the solid-fluid interfaces equates the ALE mesh deformation, $\bu_\text{m}$ in the electrolyte sub-domain to the solid deformations:
\begin{align}
    \bu_\text{m}=\bu \quad \text{on } \Gamma_\text{s-e}, \Gamma_\text{c-e}, \Gamma_\text{p-e}.
\end{align}

Values of the coefficients and material parameters appearing in the equations of Sections \ref{sec:strong form} and \ref{sec:BC} that were used in the numerical examples have been collected in Tables \ref{tab:electrode-electrolyte} and \ref{tab:rest}.

\begin{table}[]
\begin{tabular}{|c|c|c|c|c|c|}
 \hline
  \multicolumn{6}{|l|}{Material data and parameters used for the numerical examples.}\\ \hline
Symbol &  Name&Unit & LiC$_6$ & Electrolyte & NMC\\ \hline
 \multicolumn{6}{|c|}{constant}\\ \hline
$F$ & Faraday's constants  &  C/mol &\multicolumn{3}{c|} {96487}\\ \hline
$R$ & Universal gas constant & J/(mol$\cdot$K)& \multicolumn{3}{c|}{8.3143}\\ \hline
$\theta_0$& Initial temp &K& \multicolumn{3}{c|} {298} \\ \hline
 \multicolumn{6}{|c|}{Electrochem parameters}\\ \hline
$\alpha_a$ & Transfer coeff\cite{ZWJES2017} &-& 0.5 &-&0.5\\ \hline
$\alpha_c$ & Transfer coeff\cite{ZWJES2017} &-& 0.5 &-&0.5\\ \hline
$k_0$ & kinetic rate constant\cite{ZWJES2017} &$\sqrt{\text{mol}}/(\text{m}^2\text{s})$& $8.0\times 10^{2}$ & -& $8.0\times 10^{2}$\\ \hline
$\sigma$ & Electric conductivity\cite{ZWJES2017}& $\text{S}/\text{m}$& $1.5\times 10^2$& Eq.\cite{ZWJES2017} 14& $0.5\times 10^2$  \\ \hline
$D_\text{s}$ & Diffusivity of lithium\cite{ZWJES2017}& $\text{m}^2/\text{s}$& $5\times 10^{-13}$&-& $1\times 10^{-13}$\\ \hline
 $t_0^+$ & Transfer number\cite{ZWJES2017} &- &-&0.2 &-\\ \hline 
 \multicolumn{6}{|c|}{Therm parameters}\\ \hline
$ \rho$& Density\cite{ZWJES2017}& $\text{kg}/\text{m}^3$& $2.5\times 10^{3}$ & $1.1\times 10^{3}$ & $2.5\times 10^{3}$\\ \hline
$C_p$ & Specific heat\cite{ZWJES2017}& $\text{J}/(\text{kg}\cdot\text{K})$& $7\times 10^{2}$& $7\times 10^{2}$& $7\times 10^{2}$   \\ \hline
$\lambda$&Therm conductivity\cite{ZWJES2017} &$\text{W}/(\text{m}\cdot\text{K})$ & $1.04$  & $0.33$& $5$\\ \hline
$h$ & heat transfer coeff\cite{ZWJES2017} & $\text{W}/(\text{m}^2\cdot\text{K})$& 5& 5 & 5\\ \hline
$\Omega$ &Therm exp coeff\cite{ZWJES2017}& $1/\text{K}$ & $9.615\times 10^{-6}$ &-& $6.025\times 10^{-6}$\\ \hline
$\Omega_\text{s}$ &Therm exp coeff\cite{ZWJES2017} & $1/\text{K}$ & $6\times 10^{-6}$ & -& $6\times 10^{-6}$\\ \hline
 \multicolumn{6}{|c|}{Mechanics}\\ \hline
$E$& Young's modulus\cite{ZWJES2017}&GPa& 5.93 & -& 8.88 \\ \hline
$\nu$ & Poisson's ratio\cite{ZWJES2017} &-& 0.3& -& 0.3\\ \hline
$\eta$& dynamic viscosity\cite{Gu2000}&$\text{kg}/(\text{m}\cdot \text{s})$ &-&$1.0\times 10^{-3}$&-\\\hline
\end{tabular}
\caption{\color{black}Material data and parameters for the electrodes and electrolyte.}
\label{tab:electrode-electrolyte}
\end{table}

\begin{table}
\begin{tabular}{|c|c|c|c|c|c|c|} 
\hline
Symbol &Name &Unit& Al &carbon-binder\footnote{Estimated based on properties PVDF.}&sep. particles& Cu\\ \hline
$E$&Young’s modulus\cite{ShiJPS2011,Awarke2011}& Gpa&70& 2.3&0.5 &117\\\hline
$\nu$&Poisson's ratio\cite{ShiJPS2011,Awarke2011}& &0.34& 0.35& 0.35&0.35 \\\hline
$\rho$& Density\cite{ShiJPS2011}& $\text{kg}/\text{m}^{3}$&2700& 1780&1200&8900\\\hline
$\lambda$& Therm. conduct.\cite{XuEnergy2011,Jeon2011}&$\text{W}/(\text{m}\cdot \text{K})$&160&0.19&1&400 \\\hline
$C_p$&Specific heat\cite{XuEnergy2011,Jeon2011}&$ \text{J}/(\text{kg}\cdot \text{K}$ &903 &700&700&385 \\\hline
$\Omega_{\theta}$&Therm. Expan. Coeff.\cite{ShiJPS2011}& $1/\text{K}$&$23.6\times 10^{-6}$&$190\times 10^{-6}$&$13.32\times 10^{-5}$&$17\times 10^{-6}$\\\hline
$\sigma_\text{s}$& Elect. Conduct.\cite{Jeon2011}&$\text{S}/\text{m}$&$6\times10^{7}$&$100$ &&$3.8\times10^{7}$\\\hline
\end{tabular}
\caption{\color{black}Material data and parameters for the carbon-binder, separator and current collectors.}
\label{tab:rest}
\end{table}

\section{Numerical treatment}
\label{sec:Numerical}
Equations (\ref{eq:ConsCli}), (\ref{eq:ConsCli+}), (\ref{eq:reducedPhis}), (\ref{eq:heatElectrode}-\ref{eq:heatElectrolyte}), (\ref{eq:strongformElasticity}), (\ref{eq:stoke}) and (\ref{eq:continueEq}) are coupled and highly nonlinear. Furthermore, the interface conditions, (\ref{pa:irreversibleentropicheat}--\ref{pa:reversibleelectrochemicalreactionheat}), (\ref{eq:interfaceV-u}) and (\ref{eq:interfaceU-v-s}-\ref{eq:interfaceU-v-p})  and many coefficients in the partial differential equations depend on the primary variables, introducing further nonlinearity to the system of equations. Here, they are written in weak form and solved by the finite element method using code implemented in {\color{black}{\tt mechanoChemFEM}, a library developed by the authors for modeling of mechano-chemical problems. It uses the finite element method and is based on the open source finite element library {\tt deal.II}\cite{Bangerth2007}.
Our code is parallelized using MPI and the initial and boundary value problems presented here can be found at \url{https://github.com/mechanoChem/mechanoChemFEM/tree/example}.} 
\subsection{The Galerkin weak form and the finite element formulation}
For a generic, finite-dimensional field $u^h$, the problem is stated as follows: Find $u^h\in \mathscr{S}^h \subset \mathscr{S}$, where $\mathscr{S}^h= \{ u^h \in \mathscr{H}^1(\Omega_0) ~\vert  ~u^h = ~\bar{u}\; \mathrm{on}\;  \Gamma_{0}^u\}$,  such that $\forall ~w^h \in \mathscr{V}^h \subset \mathscr{V}$, where $\mathscr{V}^h= \{ w^h \in\mathscr{H}^1(\Omega_0)~\vert  ~w^h = ~0 \;\mathrm{on}\;  \Gamma_{0}^u\}$, the finite-dimensional (Galerkin) weak form of the problem is satisfied. The variations $w^h$ and trial solutions $u^h$ are defined component-wise using a finite number of basis functions,
\begin{equation}
w^h = \sum_{a=1}^{n_\mathrm{b}} c^a N^a, \quad \qquad u^h = \sum_{a=1}^{n_\mathrm{b}} d^a N^a,
\label{eq:basisdef}
\end{equation}
\noindent where $n_\mathrm{b}$ is the dimensionality of the function spaces $\mathscr{S}^h$ and $\mathscr{V}^h$, and $N^a$ represents the basis functions. To obtain the Galerkin weak forms, we multiply each strong form by the corresponding weighting function, integrate by parts and apply boundary conditions appropriately, leading to the following functionals:
\begin{align}
\mathscr{R}_{C_\text{Li}}&=\int_{\Omega_{\text{s}}}w_{c_\text{Li}}\frac{\partial C_\text{Li}}{\partial t}\text{d}v-\int_{\Omega_{\text{s}}} \nabla w_{c_\text{Li}}\bj \text{d}v+\int_{\Gamma_\text{s}}w_{c_\text{Li}}\bj\cdot\bn_\text{s} \text{d}s=\bzero,
\label{eq:WeakformCli}
\end{align}

\begin{align}
\mathscr{R}_{C_{\text{Li}^+}}&=\int_{\Omega_{\text{e}}}w_{c_{\text{Li}^+}}\frac{\partial C_{\text{Li}^+} }{\partial t}\text{d}v-\int_{\Omega_e} \nabla w_{c_{\text{Li}^+}}\bj_+ \text{d}v+\int_{\Gamma_\text{e}}w_{c_{\text{Li}^+}}\bj_+\cdot\bn_\text{e} \text{d}s=\bzero,
\label{eq:WeakformCli+}
\end{align}

\begin{align}
\mathscr{R}_{\phi_\text{S}}&=-\int_{\Omega_\text{s},\Omega_\text{c}} \nabla w_{\phi_\text{S}}\nabla\bi_\text{S}\text{d}v+\int_{\Gamma_\text{s-e}}w_{\phi_\text{S}}\bi_\text{S}\cdot \bn_\text{s}+\int_{\Gamma_\text{c}}w_{\phi_\text{S}}\bi_\text{S}\cdot \bn_\text{c} \text{d}s=\bzero,
\label{eq:WeakformPhis}
\end{align}

\begin{align}
\mathscr{R}_{\phi_\text{E}}&=-\int_{\Omega_e} \nabla w_{\phi_\text{E}}\nabla\bi_\text{E}\text{d}v+\int_{\Gamma_\text{e}}w_{\phi_\text{E}}\bi_\text{E}\cdot \bn_\text{e} \text{d}s=\bzero,
\label{eq:WeakformPhie}
\end{align}

\begin{align}
\mathscr{R}_\theta=\int_{\Omega_{\text{total}}}w_\theta\rho C_p \frac{\partial \theta}{\partial t}\text{d}v+\int_{\Omega_{\text{total}}}\nabla w_\theta \lambda \nabla \text{d}v- \int_{\Omega_{\text{total}}} w_\theta Q  \text{d}v- \int_{\Gamma_\text{outer}}w_\theta\lambda \nabla \theta\cdot \bn_\text{outer} \text{d}s = \bzero,
\label{eq:WeakformT}
\end{align}

\begin{align}
\mathscr{R}_u=\int_{\Omega_{\text{s}}\cup\Omega_{\text{c}}\cup\Omega_{\text{p}}}\nabla \bw_u\bT \text{d}v- \int_{\Gamma\text{s-e}}\bw_u \bff \cdot \bn_\text{s} \text{d}s -\int_{\Gamma\text{e-c}}\bw_u \bff \cdot \bn_\text{c} \text{d}s -\int_{\Gamma\text{e-p}}\bw_u \bff \cdot \bn_\text{p} \text{d}s= \bzero,
\label{eq:WeakformU}
\end{align}

\begin{align}
\mathscr{R}_{v}&=-\int_{\Omega_e} \nabla \bw_{v}(-2\eta\varepsilon(\bv)+p) \text{d}v+\int_{\Gamma_\text{e}}\bw_{v}(-2\eta\varepsilon(\bv)+p)\cdot \bn_\text{e} \text{d}s=\bzero,
\label{eq:WeakformV}
\end{align}

\begin{align}
\mathscr{R}_{p}&=\int_{\Omega_e}w_{p}\nabla \cdot(\rho\bv)\text{d}v=\bzero,
\label{eq:WeakformP}
\end{align}

\begin{align}
\mathscr{R}_{u_\text{m}}=\int_{\Omega_{\text{e}}}\nabla \bw_{u_\text{m}}\bT_\text{m} \text{d}v- \int_{\Gamma_\text{e}}\bw_{u_\text{m}} \bff_\text{m} \cdot \bn_\text{e} \text{d}s = \bzero,
\label{eq:WeakformUmesh}
\end{align}

The fields $w_{c_\text{Li}}$, $w_{c_{\text{Li}^+}}$, $w_{\phi_\text{S}}$, $w_{\phi_\text{E}}$, $w_\theta$, $\bw_u$, $\bw_{v}$, $w_{p}$ and $\bw_{u_\text{m}}$ are weighting functions for the corresponding primary variables. The sub-domains of active material, electrolyte, current collector and polymeric separator particles are $\Omega_{\text{s}}$, $\Omega_{\text{e}}$, $\Omega_{\text{c}}$ and $\Omega_{\text{p}}$, respectively. The boundaries of the active material and electrolyte sub-domains are $\Gamma_\text{s}$ and $\Gamma_\text{e}$, respectively. We define the domain $\Omega_{\text{total}}=\Omega_{\text{s}}\cup \Omega_{\text{e}}\cup\Omega_{\text{c}}\cup\Omega_{\text{p}}$, for the entire battery cell, and $\Gamma_\text{outer}$ as its outer boundary. All the weak forms are evaluated in the current configuration. For volume integration between the current and reference configurations we have $\text{d}v=J\text{d}V$, where $J=\det F$. 

To account for the deforming interfaces, especially the interface between active material and electrolyte (SEI), Nanson's formula is used in surface integrals. At a surface or interface element whose areas in the reference and  deformed configurations are $\text{d}S$ and $\text{d}s$, respectively, with unit outward normal $\bN$ to $\text{d}S$, we have

\begin{align}
\text{d}s=||\bF^{-T}\cdot\bN||J\text{d}S
\end{align}

Time integration is achieved by the Backward-Euler algorithm. Also, for equation (\ref{eq:interfaceV-u}), we use the discretized formula for the time derivative:
\begin{align}
\bv^{N+1}= \frac{\bu^{N+1}-\bu^{N}}{\Delta t} \quad \text{on } \Gamma_\text{s-e}
\label{eq:v_u_discrete}
\end{align}
where superscript $N$ denotes a state at time $t_N$. This condition is directly embedded as a constraint between the finite element degrees of freedom for $\bv$ and $\bu$.

{\color{black}Since our {\tt mechanoChemFEM} code is based on {\tt deal.II}, which uses hexahedral meshes, we used bilinear or trilinear basis functions for the primary variables other than the velocities for which we used biquadratic or triquadratic basis functions in order to pass the Ladyzhenskaya-Babuska-Brezzi (LBB) conditions. Meshes were generated in {\tt Cubit} and were, in general unstructured. The smallest elements were $\sim 0.1\;\mu$m in diameter at which resolution the solutions were converged. See Figure \ref{fig:ParticleModel} for dimensions of the inital and boundary value problems. A typical problem yields $\sim 10^5$ degrees of freedom for two dimensions and $\sim 10^6$ degrees of freedom for three-dimensional problems. The SuperLU direct linear solver was used with the Newton-Raphson method with line search for nonlinear solutions. A convergence tolerance of $10^{-6}$ was imposed on the Euclidean norms of the residual vectors in Equations (\ref{eq:WeakformCli}-\ref{eq:WeakformUmesh}) (after normalization to 1 initially). Note that {\tt deal.II} does not currently allow the solving of different partial differential equations over different domains, in combination with distributed meshes. Additionally, due to the coupling of many fields, the bandwidth of the linear system is of $\mathcal{O}(10^3)$. These features combine to make the problem highly memory intensive, and in particular make large three-dimensional simulations challenging.}

\subsection{Algorithmic differentiation}
For a highly non-linear set of equations, numerical differentiation is inaccurate and ultimately unstable. An effective and efficient alternative is the use of algorithmic (or automatic) differentiation (AD), which works by application of the chain rule to algebraic operations and functions (polynomial, trigonometric, logarithmic, exponential or reciprocal) in the code. AD thus works to machine precision at a computational cost that is comparable to the cost of evaluation of the original equations. We use AD in this work to linearize Equations (\ref{eq:WeakformCli}-\ref{eq:WeakformUmesh}), and compute the Jacobian matrix. Specifically, we use the {\tt Sacado} package, which is part of the open-source {\tt Trilinos} project\cite{Sacado2005,sacado2012}.

\section{Numerical Examples}
\label{sec:Results}
\begin{figure}[hbtp] 
\centering
\includegraphics[scale=0.25]{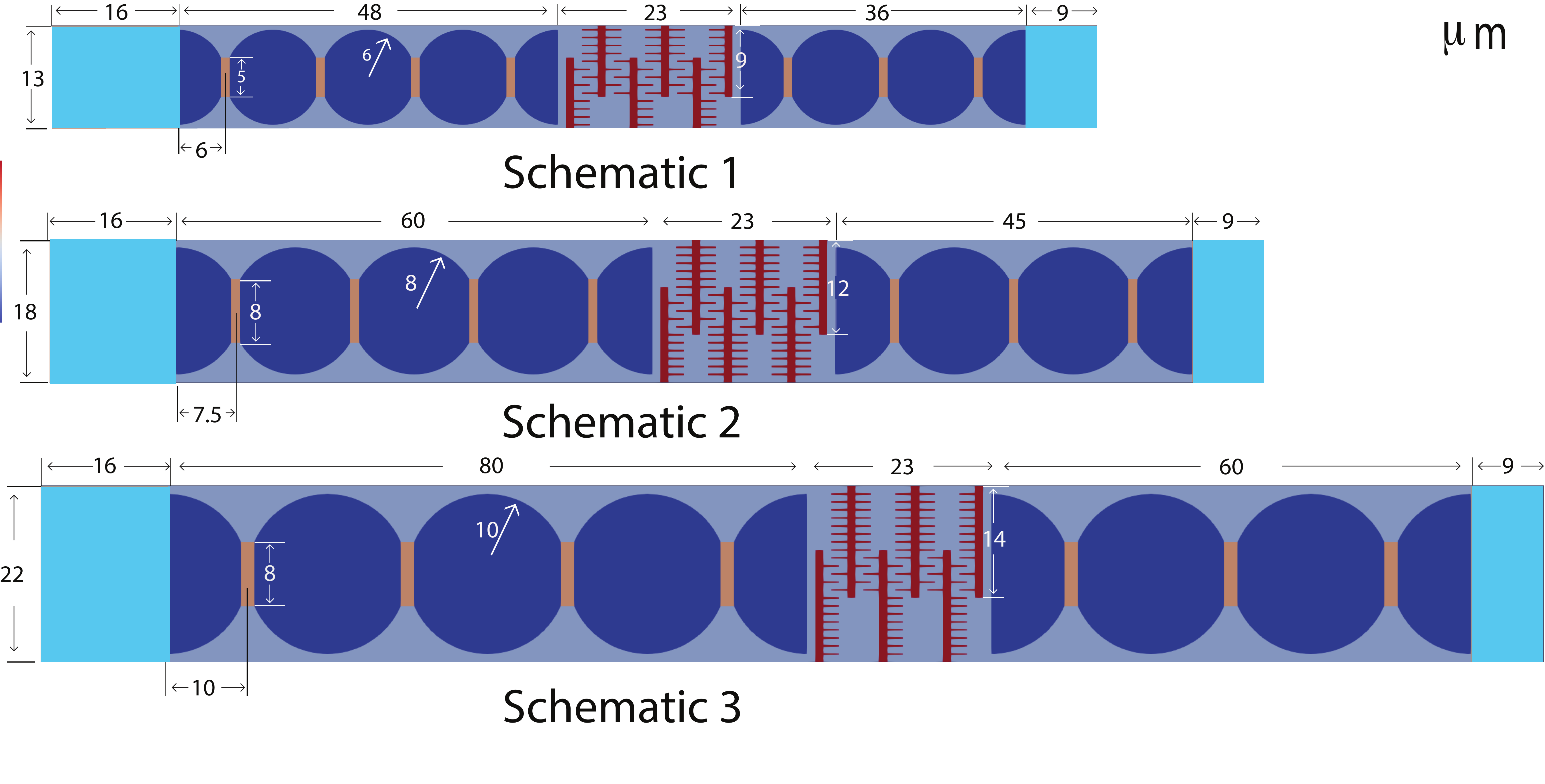}
\caption{{\color{black}Schematics showing the dimensions of the initial/boundary value problem. In all three cases, the volume fractions for active particles and carbon-binder are $\sim 0.7$ and $\sim 0.03$, respectively, and the porosity is $\sim 0.27$. {\color{black}The specific areas of the active particle-electrolyte are $0.25\;\mu\text{m}^{-1}$, $0.18\;\mu\text{m}^{-1}$, and $0.15\;\mu\text{m}^{-1}$ from top to bottom.} The thickness of the carbon-binder is $\sim 1\;\mu$m. For the polymeric separator, the thickness of the main branch is $\sim 1\;\mu$m, and $\sim 0.2\; \mu$m for the sub-branches, which is comparable to the microstructure seen in SEM images.\cite{wangSR2016}}}
\label{fig:ParticleModel} 
\end{figure}

For our particle scale model, the three-dimensional reconstruction of porous electrodes could be achieved by X-ray tomography, using one of many published methods.\cite{Thiele2015} Here, for simplicity, we model the active material as a sphere in three dimensions and circle in two dimensions. {\color{black}The assumption of spherical symmetry is well-suited to some electrode materials such as Nickel Manganese Cobalt (NMC) oxide, although others, such as natural graphite, exhibit more elongated particles for which this assumption is less appropriate. } The particles are connected by carbon-binder providing a path for current flow, {\color{black}while preventing the lithium from diffusing across active particles.} Motivated by the multi-layered microstructure of the separator\cite{wangSR2016}, {\color{black}we model it with six layers of elongated polymeric particles with sub-branches }. The cell is sandwiched between two current collectors. 
{\color{black}We consider an idealized geometry with particles aligned along the electrode thickness. Such a geometry is not representative of an actual electrode material as it implies periodic variation of microstructural properties (such as porosity) along the electrode thickness (the $\be_1$-direction in Figure \ref{fig:schematicParticleModel}), which is not borne out by tomography-based microstructural analysis. However, this simple geometry matches the representation of electrode materials portrayed in many macroscopic models and is adequate for a first demonstration of the capabilities of our model.} {\color{black}In these geometries, the carbon-binder is found between active
particles, in contrast with treatments such as that by Trembacki et. al.\cite{TrembackiJES2017}, who represented the binder as being a uniform coating on the surface of active particles near the particle contacts but not separating active particles. The carbon-binder thickness is $\sim 1\;\mu$m, which is a somewhat higher than the more representative $\sim 100$ nm.} Our computational framework is in three dimensions, but for ease of interpretation of the coupled physics, and to avoid computational complexity in this first communication, the model is mainly presented in two dimensions. {\color{black} However, we also demonstrate the fully three-dimensional case in the Appendix}. We consider three microstructures with different particle sizes. The schematic in Figure \ref{fig:ParticleModel} shows the dimensions of the domain and volume fractions of the different materials.

\begin{figure}[hbtp]
\centering
\includegraphics[scale=0.3]{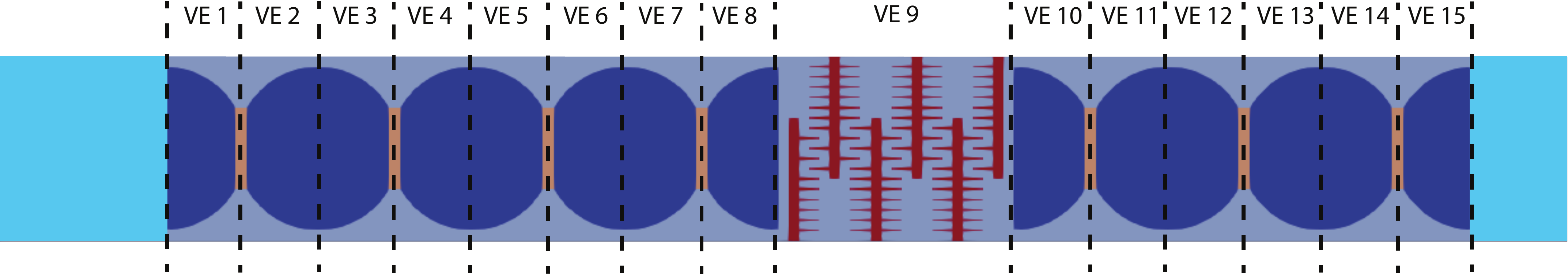}
\caption{\color{black}Volume elements defined over the domain.}
\label{fig:RVE} 
\end{figure}

{\color{black}To study the volume change driven primarily by lithium intercalation, we define eight volume elements over the negative electrode, six over the positive electrode and one for the separator, as shown in Figure \ref{fig:RVE}.}
For each volume element, we can evaluate the total volume and volume for each component:
\begin{align}
V_\text{total}=V_\text{s}+V_\text{e}+V_\text{b}
\label{eq:V_total}
\end{align}
where $V_\text{s}$ is the volume of the active material, $V_\text{e}$ is the volume of the electrolyte and $V_\text{b}$ is the volume of the carbon-binder. The porosity of each {\color{black}volume element} is defined as:
\begin{align}
\epsilon^\text{p} =\frac{V_\text{e}}{V_\text{total}}
\label{eq:epl}
\end{align}
{\color{black}And the specific area of the active particle-electrolyte and active particle-carbon binder interfaces is defined as:
\begin{align}
a_\text{s-e}=\frac{S_\text{s-e}}{V_\text{s}}
\label{eq:a_e}\\
a_\text{s-p}=\frac{S_\text{s-b}}{V_\text{s}}
\label{eq:a_b}
\end{align}
where $S_\text{s-e}$ is interfacial area between the active material and electrolyte, and $S_\text{s-b}$ is interfacial area between the active material and carbon-binder. In two-dimensions the volume and surface area in the above formulas reduce to the surface area and curve length, respectively.}

%
{\color{black}
\subsection{The initial and boundary value problems}
The computational domain that we consider is far from the cell's boundaries in the $\be_1$ direction. We apply boundary conditions on Surfaces 3 and 4, perpendicular to $\be_1$ (see Figure \ref{fig:schematicParticleModel}) to model the far-field conditions. We firstly assume that the normal velocity of the electrolyte on Surfaces 3 and 4 vanishes, meaning no electrolyte can flow in and out. The polymeric base could also be subject to displacement boundary conditions, $\bu = \bu_\text{ext}$ imposed by other cell components, or to traction (pressure) boundary conditions: $\bT\bn=\bT_\text{ext}$ from the surrounding electrolyte. We consider, Case 1: fixed displacement, i.e. $\bu = \bzero$, and Case 2: traction-free, $\bT\bn=0$ on Surfaces 3 and 4. The motivation for this choice is that these cases represent two extremes among the possible states that can otherwise only be represented by fully resolving the structure in the $\be_1$-direction. 
We first compute the discharging process for the two cases with different far-field boundary conditions at a 1C\footnote{A 1~C rate means that a battery rated at N Ah should supply N Amperes for 1 hour.} rate, using Schematic 2 from Figure \ref{fig:ParticleModel} with a particle size of $8\;\mu$m.  A detailed computational study of  solid-fluid interactions is persented. Next, the effects of porosity evolution are studied by considering different magnitudes of intercalation strain. The discharging process for Schematics 1 and 3 are then simulated to study the effects of initial particle size.

\subsection{Results for different far-field boundary conditions}
During discharge, the temperature increases by about  35 K as shown in Figure \ref{fig:temperature}. For the given thermal expansion coefficients, the strain induced by temperature change is small compared to that due to intercalation. However, the time-varying temperature can significantly affect the coefficients of the electro-chemical equations; i.e., the transport properties, conductivity and reaction parameters as discussed in Wang et. al.\cite{ZWJES2017}. 
\begin{figure}[hbtp]
\centering
\includegraphics[scale=0.2]{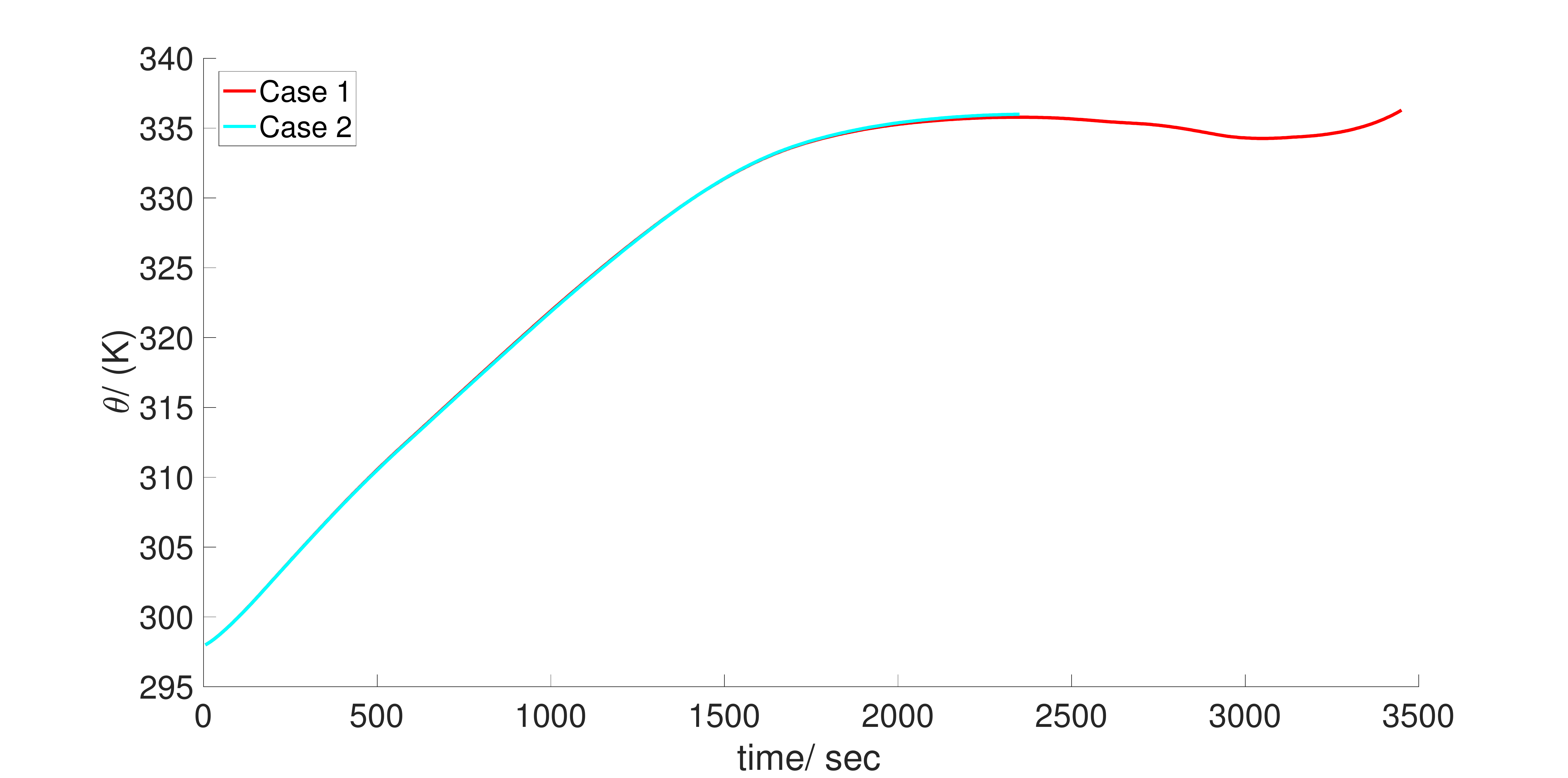}
\caption{\color{black}Temperature profile during discharge for Cases 1 and 2.}
\label{fig:temperature} 
\end{figure}

As the battery discharges, lithium de-intercalates from the negative electrode and intercalates into the positive electrode, causing active material contraction in the negative electrode and expansion in the positive electrode. The velocity of the electrolyte driven by the deforming active particles (shown in Figure \ref{fig:Pa_v}) with different far-field boundary conditions is small, such that the Peclet number $\text{Pe}\sim 1.0\times 10^{-8}$ for heat transfer and $\text{Pe}\sim 1.0\times 10^{-5}$ for ionic transport in the electrolyte. Therefore, the effects of electrolyte velocity are insignificant compared to diffusive heat and ionic transport. The contracting active particles in the negative electrode create negative pressure while the expanding active particles in the positive electrode produce positive pressure as shown in Figure \ref{fig:Pa_p}. The free traction boundary condition on the top surface (Surface 2) releases the pressure in both cases, leading to insignificant deformation of solid components due to the pressure. However the different far-field boundary conditions have significant effects on the rigid body motion of the solid components of the cell. Driven by the electrolyte flow, the  polymeric separator deforms toward the negative electrode. For Case 2, the traction free boundary condition allows the polymeric separator to bend, leading to a very narrow gap between the separator particles and active particles as shown in Figure \ref{fig:V_zoom}. Note that the mesh is highly distorted in this region and causes the simulation to diverge. Re-meshing is needed to advance the solution any further.
\begin{figure}[hbtp]
\centering
\includegraphics[scale=0.1]{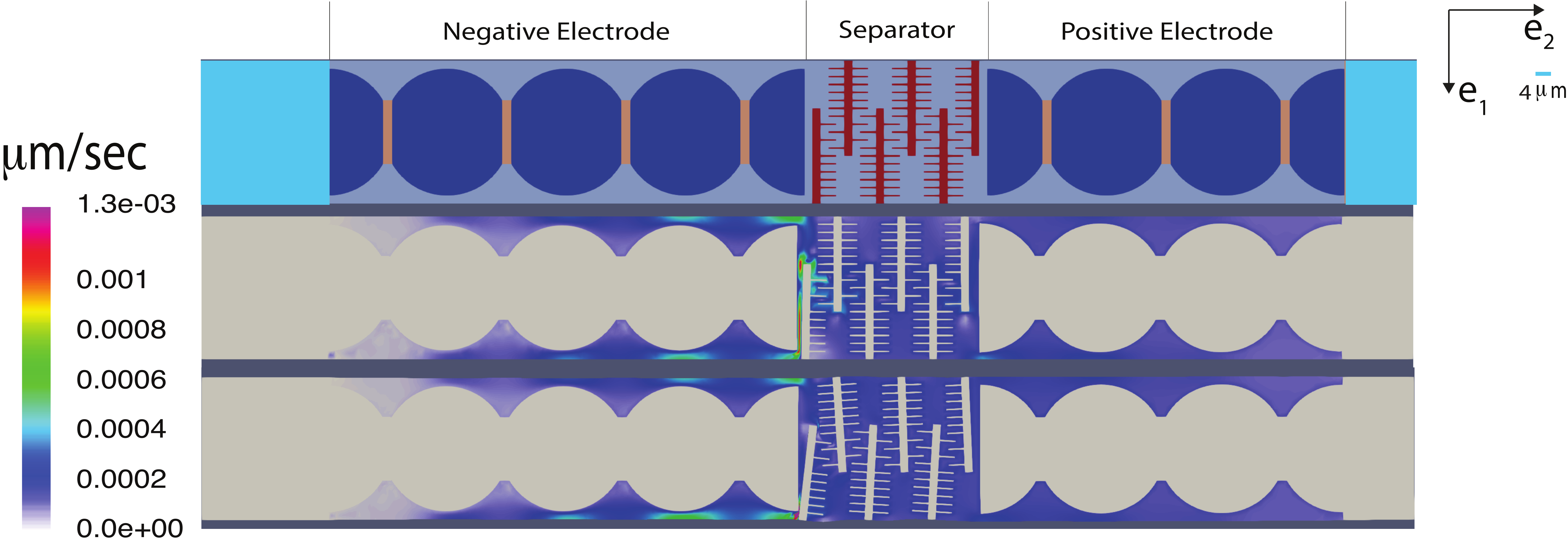}
\caption{\color{black}The magnitude of electrolyte velocity at SOC=0.5. From top to bottom: the reference state, Case 1 and Case 2.}
\label{fig:Pa_v} 
\end{figure}

\begin{figure}[hbtp]
\centering
\includegraphics[scale=0.1]{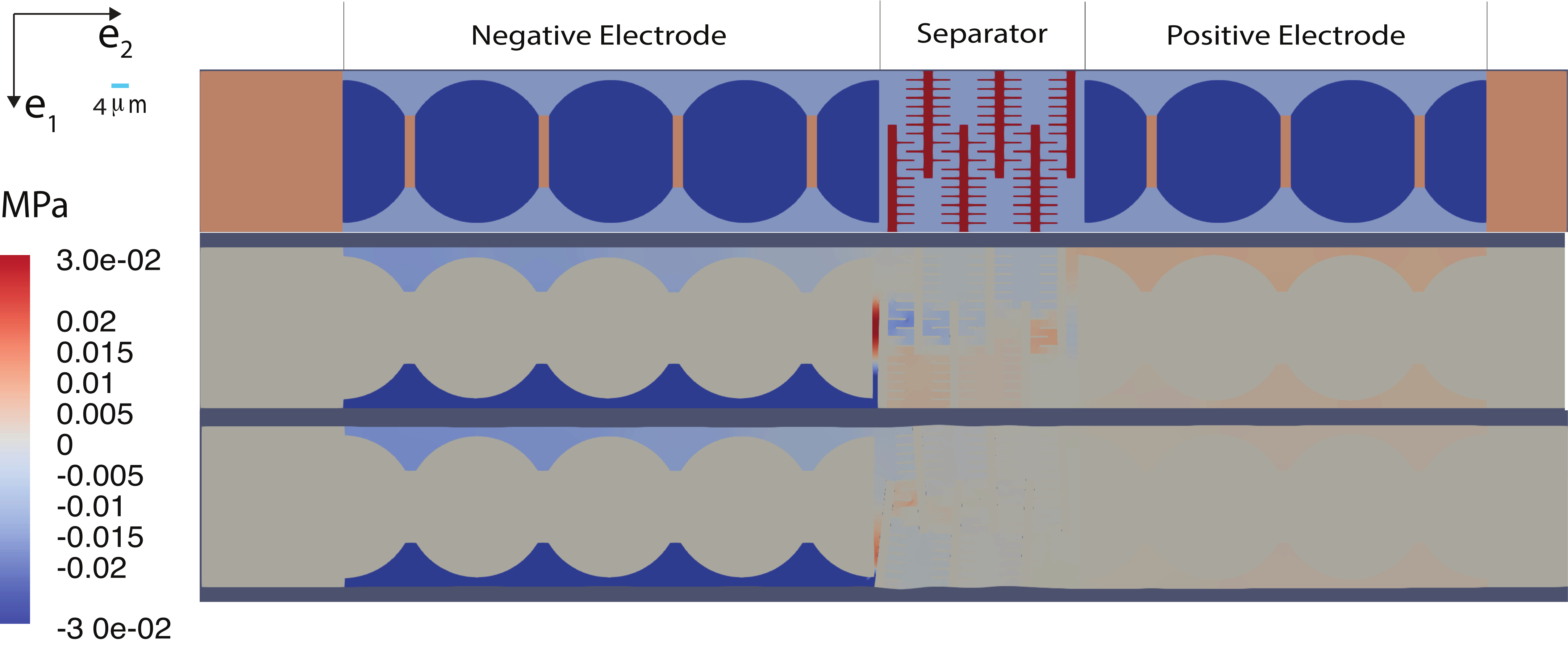}
\caption{\color{black}Pressure distribution in the fluid electrolyte at SOC=0.5. From top to bottom: reference state, Case 1 and Case 2.}
\label{fig:Pa_p} 
\end{figure}

\begin{figure}[hbtp]
\centering
\includegraphics[scale=0.1]{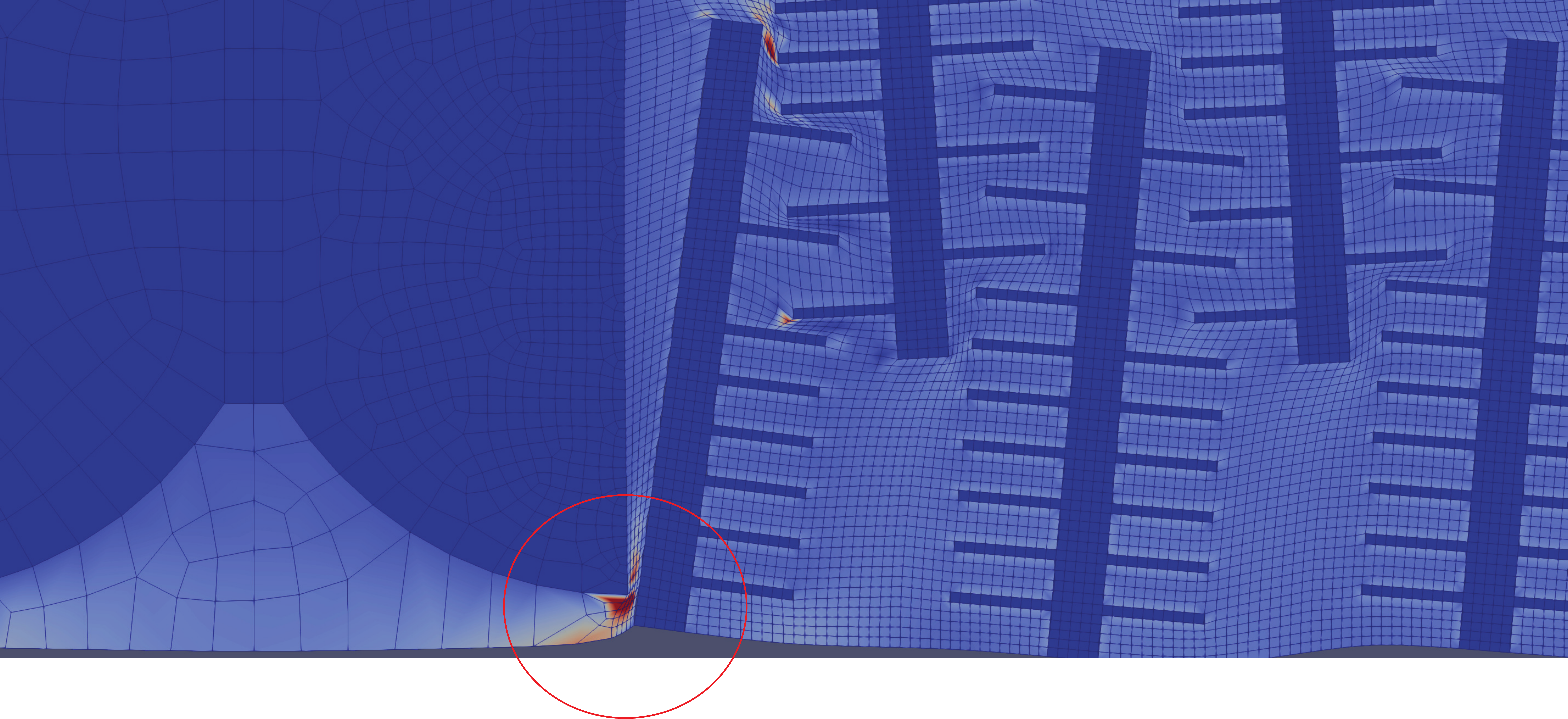}
\caption{\color{black}Expansion of the active particles and distortion of the polymeric separator particles leads to narrowing of transport paths for ions through the electrolyte. The numerical solution procedure also diverges because of extreme distortion of the ALE mesh (red circle).}
\label{fig:V_zoom} 
\end{figure}
Due to the low viscosity and pressure, there is a very low state of stress in the polymeric separator. The deformation caused by lithiation creates higher stress at the interface between active particles and the carbon-binder, and current collectors as shown in Figure \ref{fig:Pa_vonMises}.
\begin{figure}[hbtp]
\centering
\includegraphics[scale=0.1]{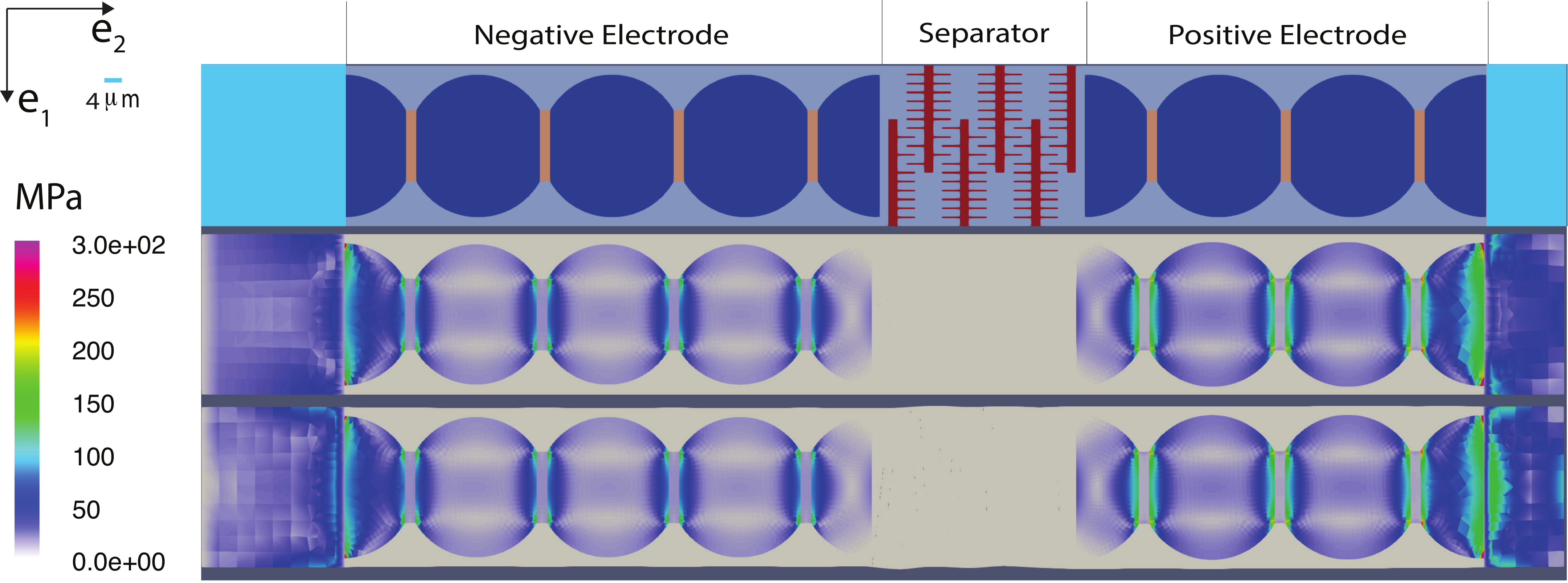}
\caption{\color{black}von Mises stress distribution in the solid components at SOC=0.5. From top to bottom: reference state, Case 1 and Case 2.}
\label{fig:Pa_vonMises} 
\end{figure}

To further analyze the effects of  far-field boundary conditions, we computed the volume change of the active material, carbon-binder and electrolyte for each volume element defined in Figure \ref{fig:RVE}. The total volume and porosity have been defined in equation (\ref{eq:V_total}) and (\ref{eq:epl}). Figure \ref{fig:epl} shows the porosity change in each volume element over time. Again, during discharge, de-intercalation of lithium causes active material contraction over the negative electrode, and the intercalation in the positive electrode causes expansion of the active material. Consequently, porosity increases over the negative electrode and decreases over the positive electrode. However, for Case 1 which corresponds to fixed displacement boundary conditions, porosity changes much more dramatically.  This is because, under fixed displacement boundary conditions, as active particles contract over the negative electrode and expand over positive electrode, the electrolyte flows from the positive to negative electrode. The volume (and mass, because of incompressibility) of electrolyte increases over the negative electrode and decreases over the positive electrode. In contrast, the traction-free boundary condition of Case 2 allows deformation of Surfaces 3 and 4 to accommodate the volume change caused by deformation of the active particles. This leads to the opposite trend for volume change of the electrolyte as shown in Figure \ref{fig:V_e}. For the same volume change of active particles, different far-field boundary conditions thus yield different trends for the change in porosity. This suggests that the porosity may change more dramatically when the whole battery cell is constrained in the $\be_1$-direction.
\begin{figure}[hbtp]
\centering
\includegraphics[scale=0.3]{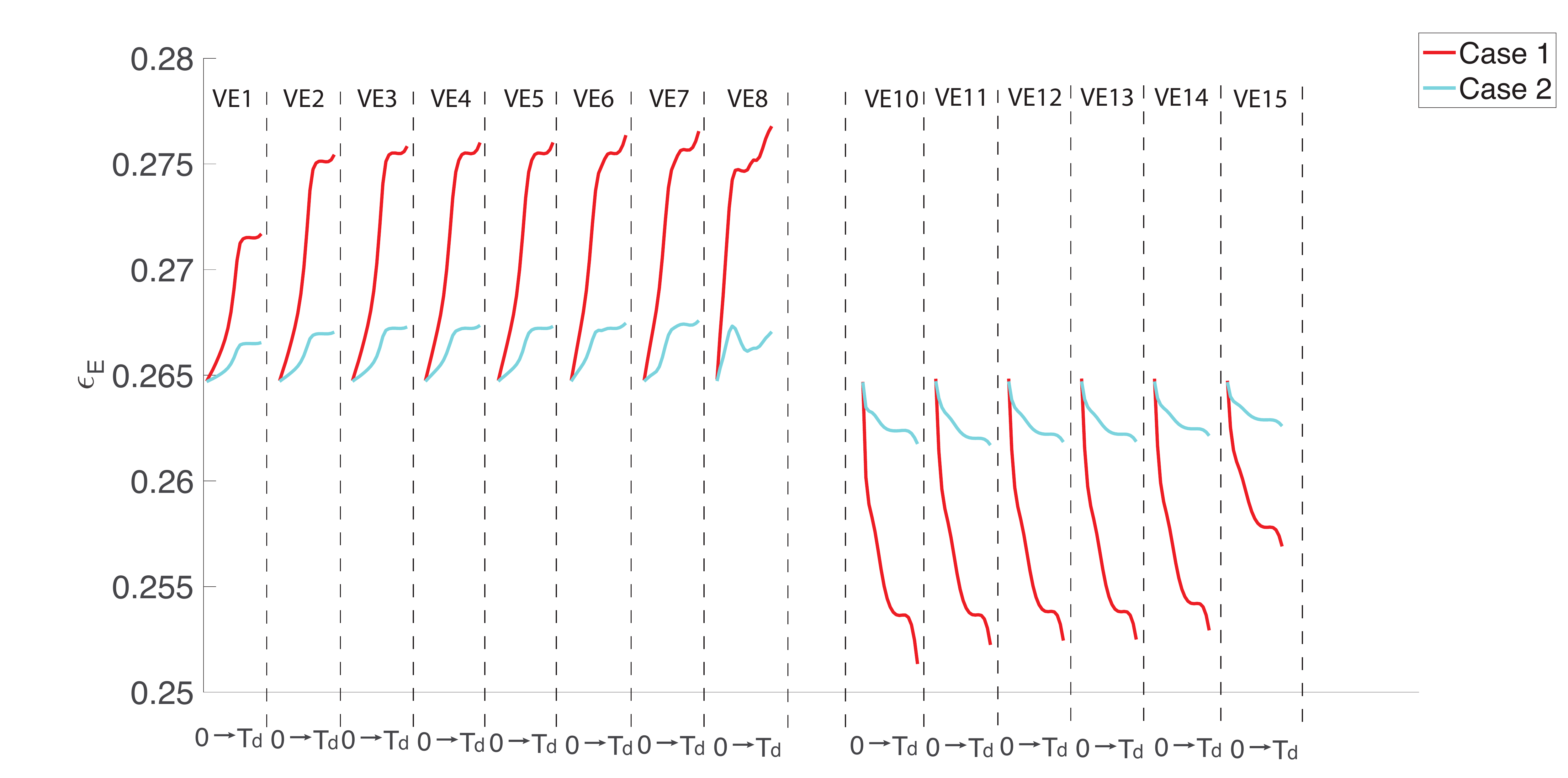}
\caption{\color{black}Porosity evolution for each volume element in the electrodes. The time interval $0 \to \text{T}_\text{d}$ denotes the duration of the discharging process.}
\label{fig:epl} 
\end{figure}

\begin{figure}[hbtp]
\centering
\includegraphics[scale=0.3]{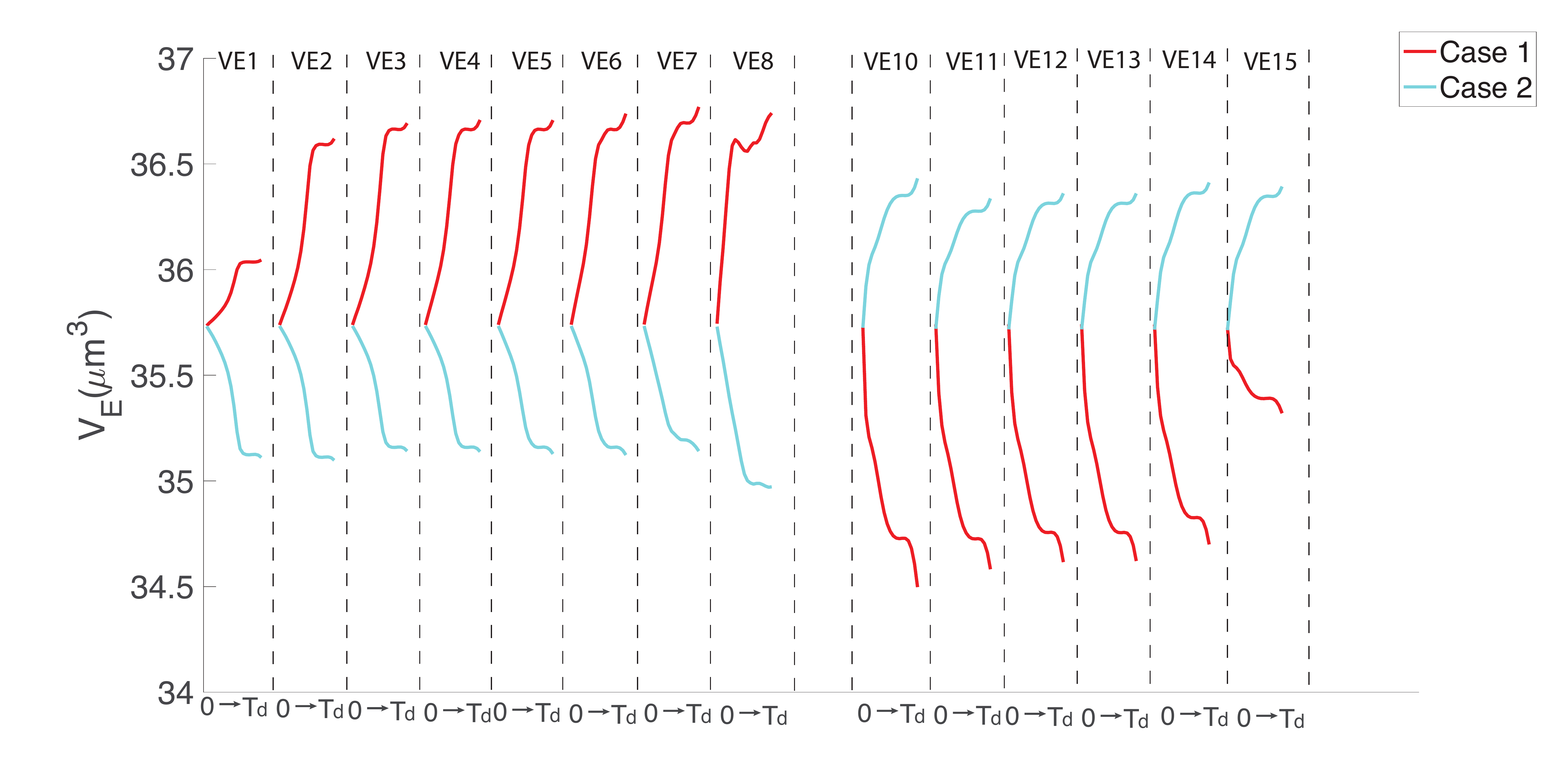}
\caption{\color{black}Evolution of the volume of electrolyte for each volume element in the electrodes. The time interval $0 \to \text{T}_\text{d}$ denotes the duration of the discharging process.}
\label{fig:V_e} 
\end{figure}

The volume change of active particles in both electrodes and its interaction with the electrolyte cause the battery cell to deform nonuniformly as shown in Figure \ref{fig:U}. Since in the configuration considered, there are more active particles in the negative electrode, the battery cell shrinks by about $0.8\%$.
\begin{figure}[hbtp]
\centering
\includegraphics[scale=0.2]{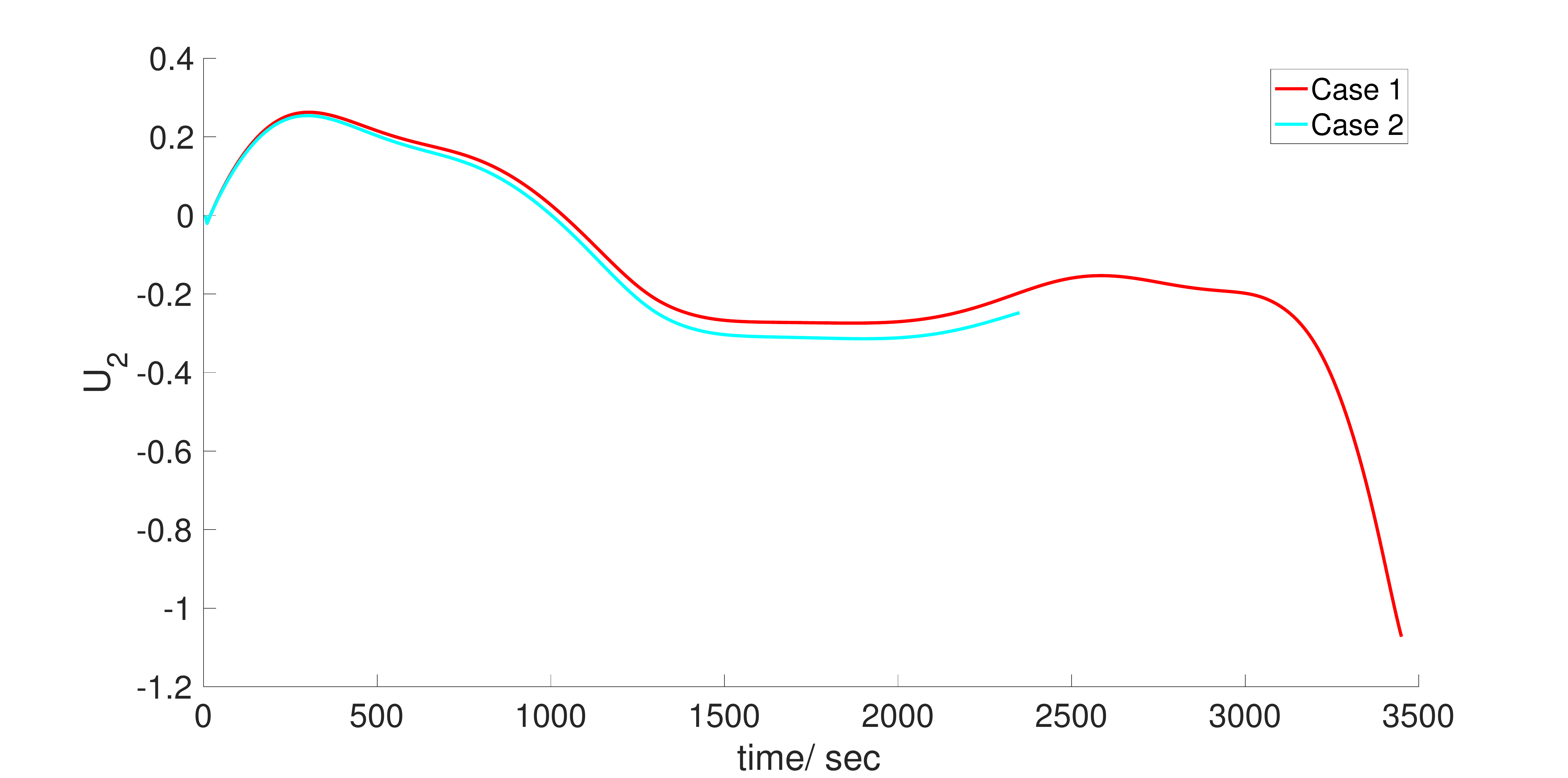}
\caption{\color{black}Displacement of Surface 2. }
\label{fig:U} 
\end{figure}

As shown in Figure \ref{fig:Pa_cli}, the distribution of lithium is non-uniform along the cell and within each particle. The ratio of interface area between active material and electrolyte to volume of active material is larger for the half-particles located closest to the separator, and the lithiation process consequently is faster in them, leading  to more non-uniform distributions. While this large interface area is entirely due to our choice to include half-particles adjacent to the separator, it has the merit of introducing some non-uniformity in particle shape within the cell. The Li distribution in the half-particles is also strongly affected by the Li$^+$ transport around the unsymmetric structure of the polymeric separator, resulting in unsymmetric distributions in the $e_1$-direction. 
\begin{figure}[hbtp]
\centering
\includegraphics[scale=0.1]{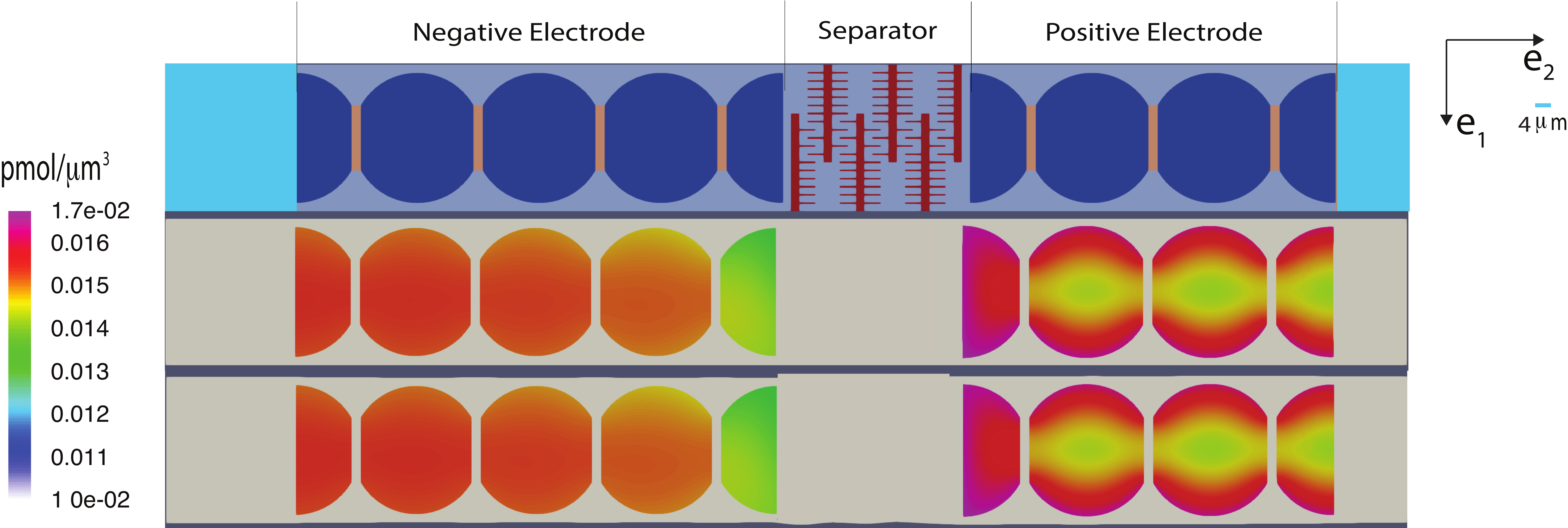}
\caption{\color{black}Lithium concentration in the deformed configuration at SOC=0.5. From top to bottom: reference state, Case 1 and Case 2.}
\label{fig:Pa_cli} 
\end{figure}
The association of lithium at the positive electrode consumes lithium ions that are produced at the negative electrode. As shown in Figure \ref{fig:Pa_cli+}, the lithium ions also are distributed non-uniformly along the cell. The
unsymmetric structure of the polymeric separator in the $\be_1$-direction also leads to an unsymmetric distribution of lithium ions along this direction near the separator.
\begin{figure}[hbtp]
\centering
\includegraphics[scale=0.1]{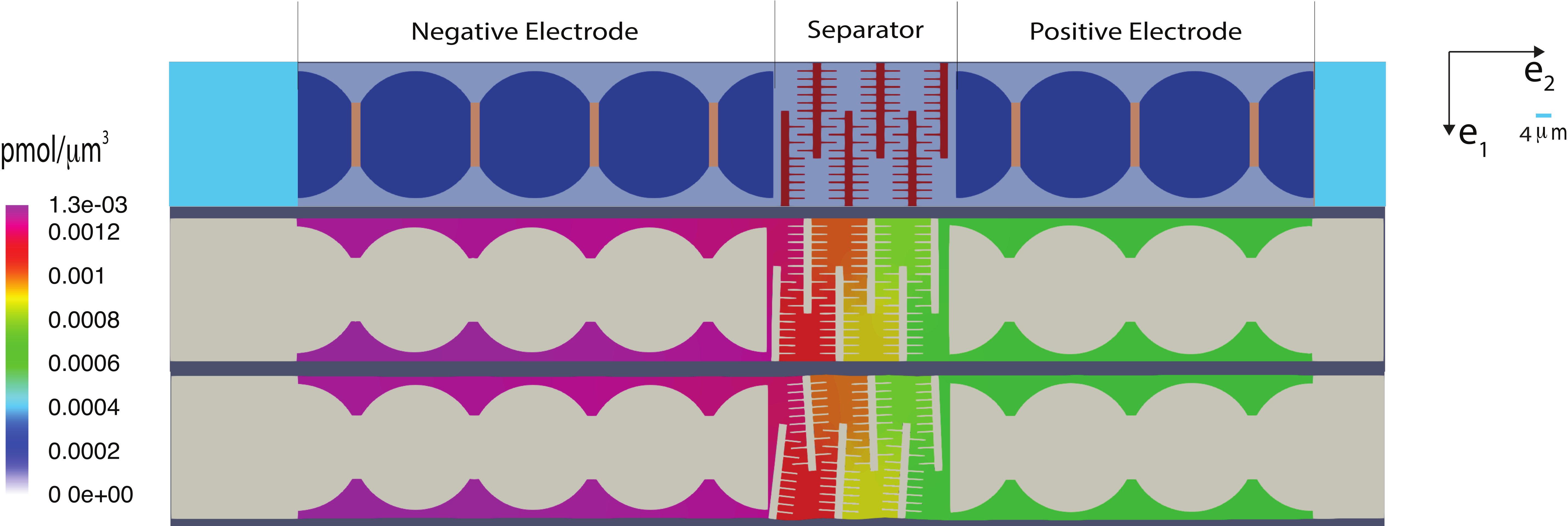}
\caption{\color{black}Lithium ion concentration in the deformed configuration at SOC=0.5. From top to bottom: reference state, Case 1 and Case 2.}
\label{fig:Pa_cli+} 
\end{figure}

While different far-field boundary conditions affect the deformation of solid components and therefore affect the porosity change, it has little effect on the electro-chemical response in total. The voltage profile (shown in Figure \ref{fig:phis}) for both cases is nearly the same.
\begin{figure}[hbtp]
\centering
\includegraphics[scale=0.2]{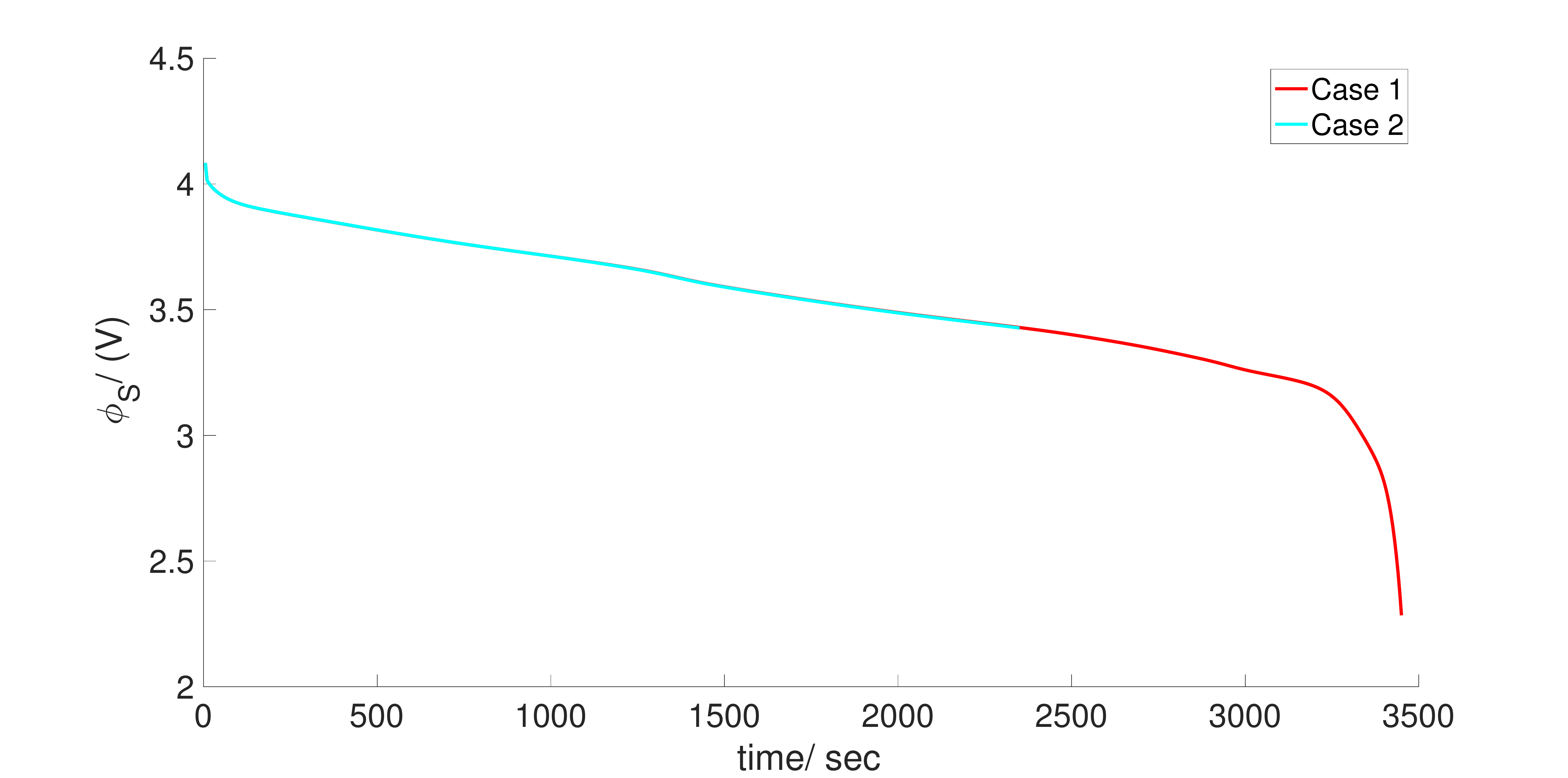}
\caption{\color{black}Electric potential,  at Surface 2 decreases during discharge. This potential is also the voltage of the cell, because $\phi_\text{S} = 0$ at Surface 1 as the reference potential. }
\label{fig:phis} 
\end{figure}

\subsection{Results for evolving porosity in materials with different intercalation strains}
The intercalation strain induced by lithiation is modeled as isotropic as shown in Equation (\ref{eq:deformFc}). The work of Wang et al. \cite{ZWJES2017} reveals that if the intercalation strain defined by the function, $\beta^c(C_\text{Li})$, remains small, there is an insignificant effect on the battery's performance as measured in its voltage response. We therefore study how  porosity evolution during battery operation affects the voltage response for a range of progressively stronger intercalation responses by scaling $\beta^c(C_\text{Li})$. The function $\beta^c(C_\text{Li})$ used by Wang et al. \cite{ZWJES2017} is denoted by $\beta^\ast(C_\text{Li})$. We then compute a discharge semi-cycle with evolving porosity using $\beta^c(C_\text{Li})=0, \beta^\ast(C_\text{Li}),2\beta^\ast(C_\text{Li})$ at a 1C current rate. The boundary conditions correspond to Case 1 which has fixed displacement boundary conditions on Surfaces 3 and 4.

Theoretically if the spherical active particles deform uniformly and maintain their original shape, the specific area (Equation (\ref{eq:a_e}) and (\ref{eq:a_b}) ) $a\sim \frac{1}{R}$. However the nonuniform, multi-dimensional nature of the governing partial differential equations causes nonuniform deformation of active particles, which do not retain their spherical shape. In our simulations the contracting active particles ``flatten'' at the active particle-electrolyte interface in the negative electrode, while the expanding active particles cause protrusion of the interface in the positive electrode. As a result, the specific area of the active particle-electrolyte interface decreases in the negative electrode and increases in the positive electrode in a nonuniform manner as shown in Figure \ref{fig:interface_beta}, while the specific area of the active particle-carbon binder interface presents the opposite trends.
\begin{figure}[hbtp]
\centering
\includegraphics[scale=0.2]{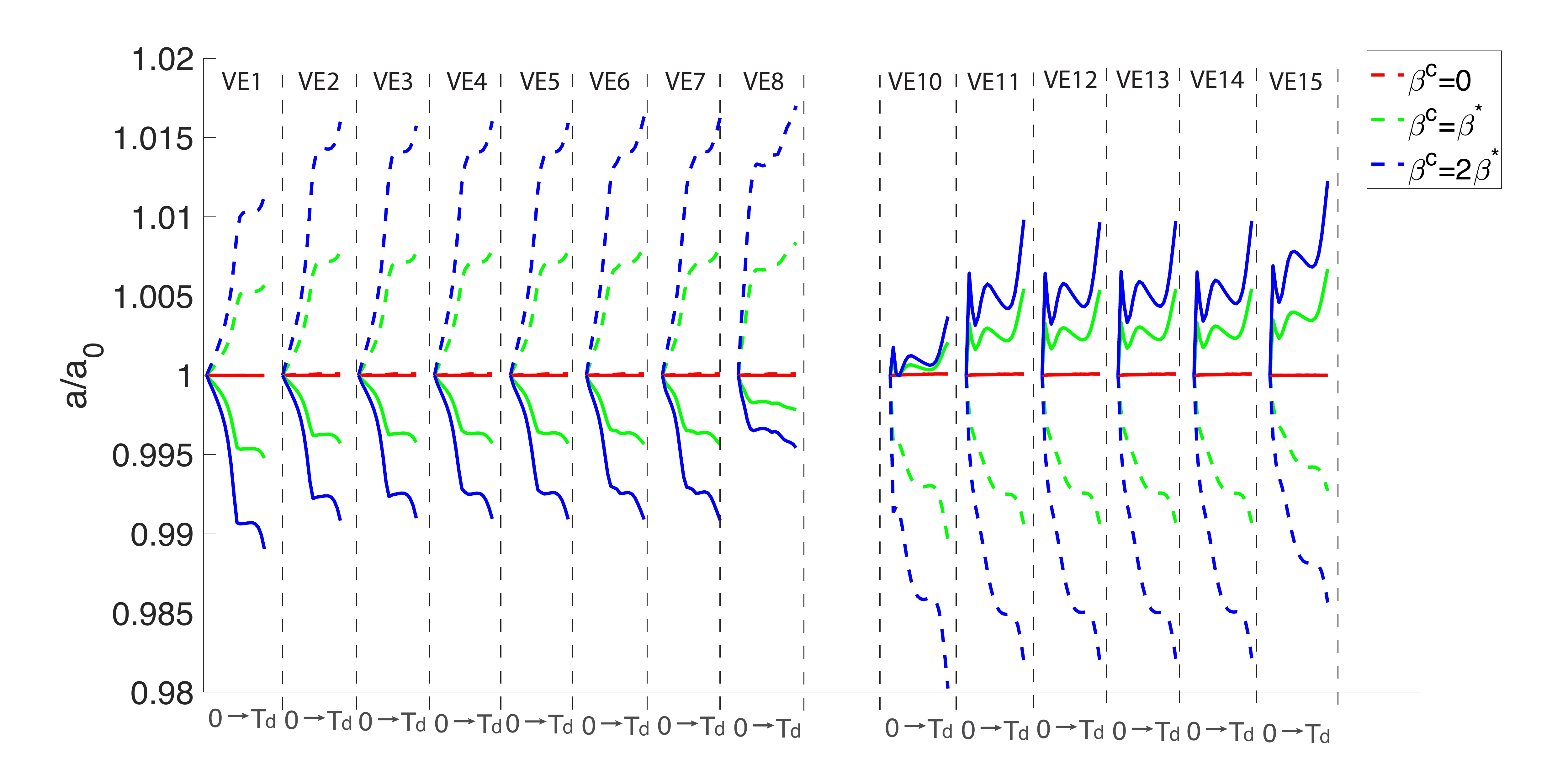}
\caption{\color{black}Normalized specific area relative to its initial value in each volume element. The dashed line denotes the active particle-carbon binder interface, the solid line denotes the active particle-electrolyte interface.}
\label{fig:interface_beta} 
\end{figure}
With stronger scaling of $\beta^c(C_\text{Li})$, the specific area changes more sharply. Lithium also gets distributed more non-uniformly among the active particles as shown in Figure \ref{fig:C_li_beta}. 
\begin{figure}[hbtp]
\centering
\includegraphics[scale=0.1]{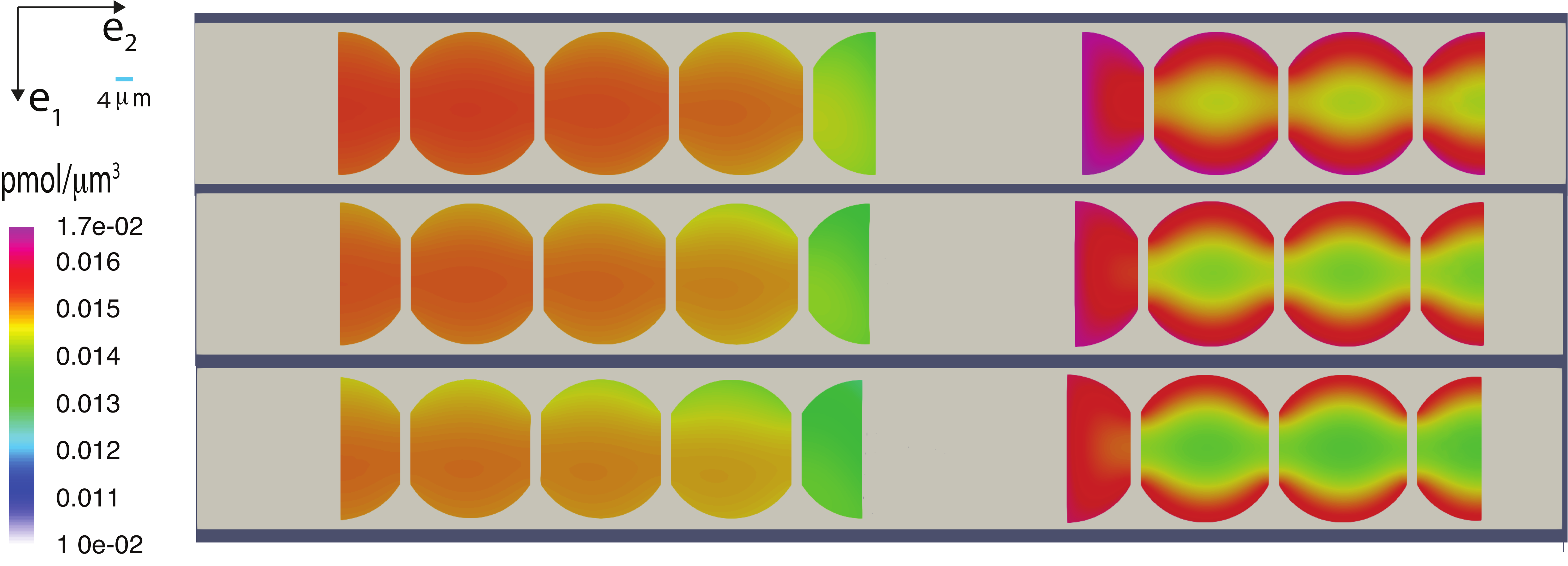}
\caption{\color{black}Lithium concentration in the deformed configuration at SOC=0.5.  From top to bottom: $\beta^c(C_\text{Li})=0,\beta^\ast(C_\text{Li}),2\beta^\ast(C_\text{Li})$, respectively.}
\label{fig:C_li_beta} 
\end{figure}
The scaling of intercalation strain has insignificant effects on the lithium ion distribution as shown in Figure \ref{fig:C_li_plus_beta}. However, the stronger scaling of $\beta^c(C_\text{Li})$ may cause large deformation of the polymeric separator and causes the mesh to be highly distorted as shown in Figure \ref{fig:V_zoom}. The computation terminates earlier than the cases with weaker scaling. 
\begin{figure}[hbtp]
\centering
\includegraphics[scale=0.1]{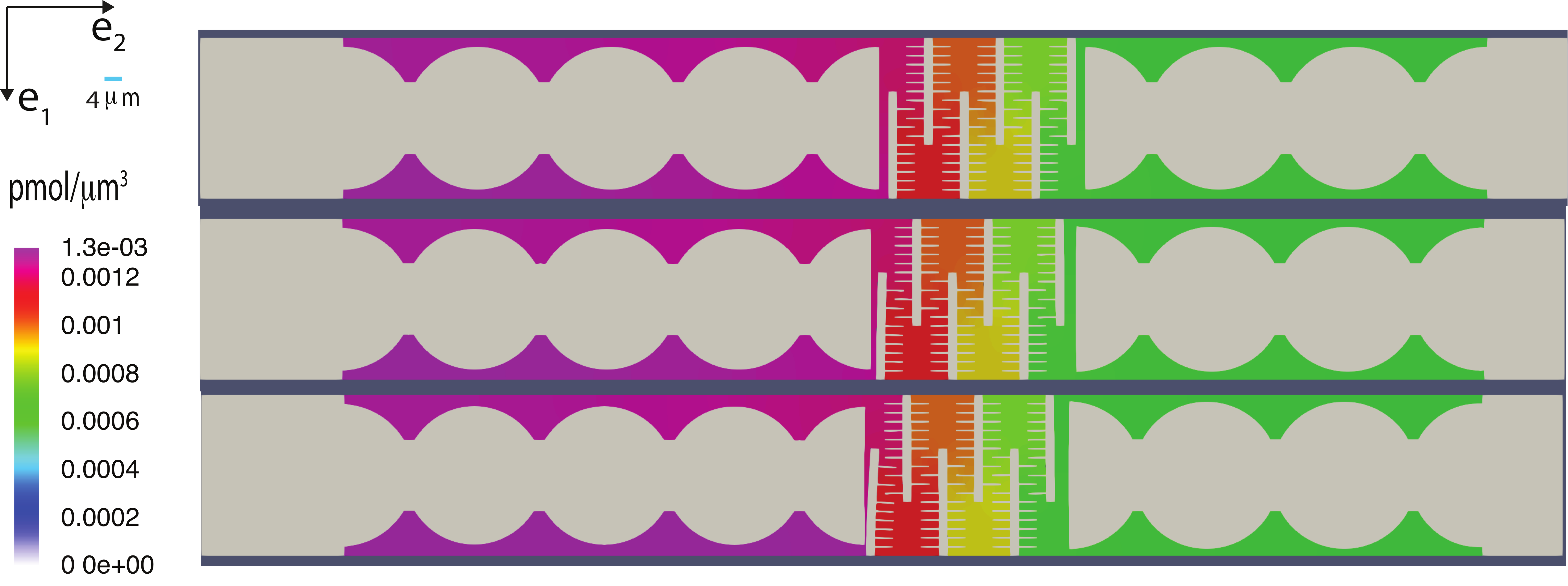}
\caption{\color{black}Lithium ion concentration in the deformed configuration at SOC=0.5.  From top to bottom: $\beta^c(C_\text{Li})=0,\beta^\ast(C_\text{Li}),2\beta^\ast(C_\text{Li})$, respectively.}
\label{fig:C_li_plus_beta} 
\end{figure} 
Table 3 
shows the thickness changes for the negative electrode, separator and positive electrode. For the case without intercalation strain, each part of the battery cell only expands slightly due to thermal expansion. For cases with intercalation strain, the thickness of the electrode changes due to the volume change of active material, while the separator shrinks slightly. 
\begin{table}
\label{ta:Thickness}
\begin{center}
\begin{tabular}{|c|c|c|c| }
\hline
 & $\beta^c=0$ & $\beta^c=\beta^*$& $\beta^c=2\beta^*$ \\\hline
negative electrode& +0.0123\% &-1.7067\%&-3.64\%\\ \hline
separator& +0.1057\% &-0.13\%&-0.4348\%\\ \hline
positive electrode& +0.0127\%  &+1.9444\% &+3.6556\% \\ \hline
\end{tabular}
\caption{\color{black}Thickness change with different intercalation strain at SOC=0.3}
\end{center}
\end{table}
As expected, the volume of active material changes more sharply with stronger scaling of $\beta^c(C_\text{Li})$, leading to larger changes in porosity as shown in Figure \ref{fig:epl_beta}.  Also, the voltage drops more rapidly as shown in Figure \ref{fig:phis_beta}.
\begin{figure}[hbtp]
\centering
\includegraphics[scale=0.2]{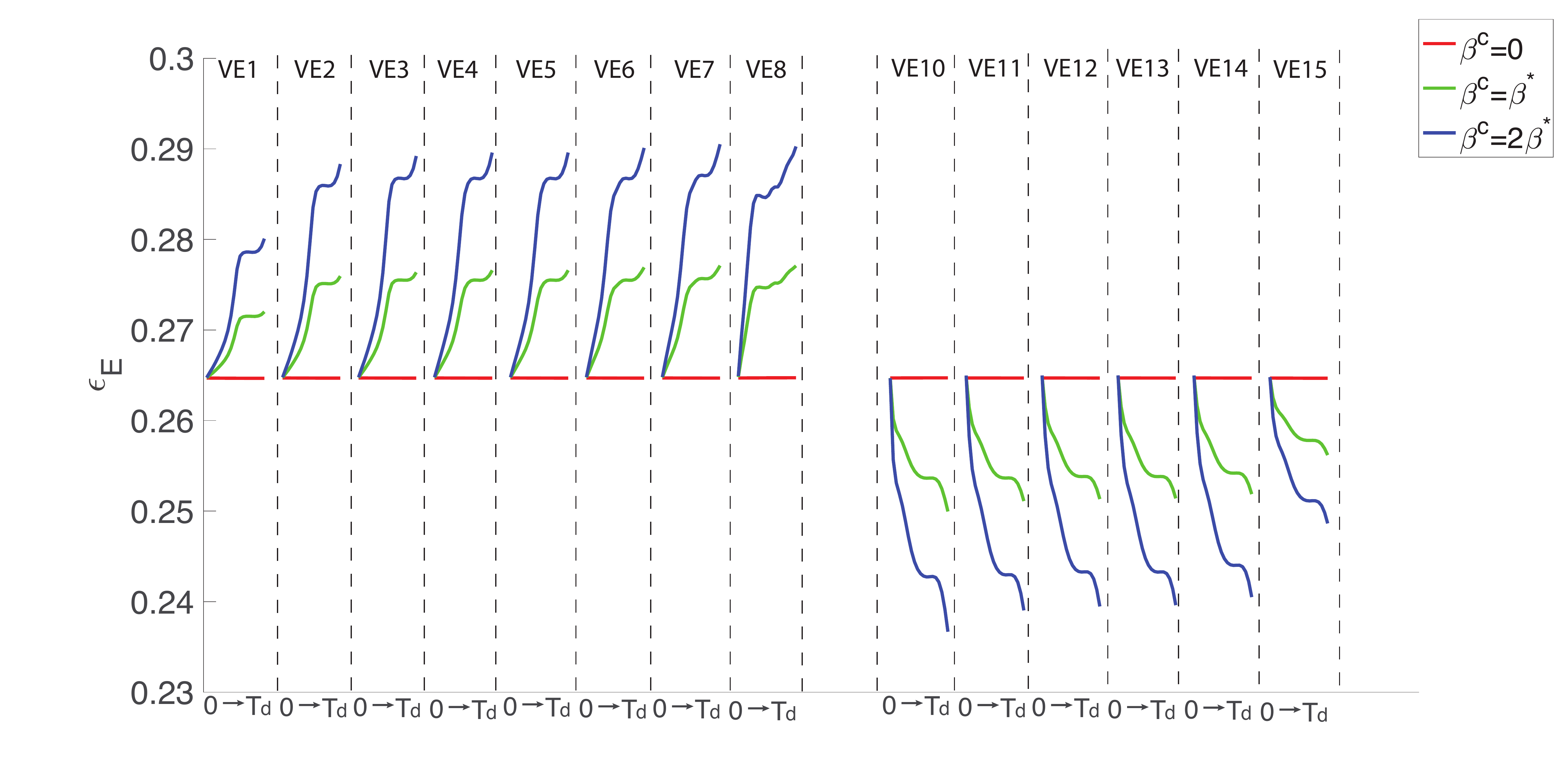}
\caption{\color{black}Porosity evolution in each volume element in the electrode.}
\label{fig:epl_beta} 
\end{figure}

\begin{figure}[hbtp]
\centering
\includegraphics[scale=0.2]{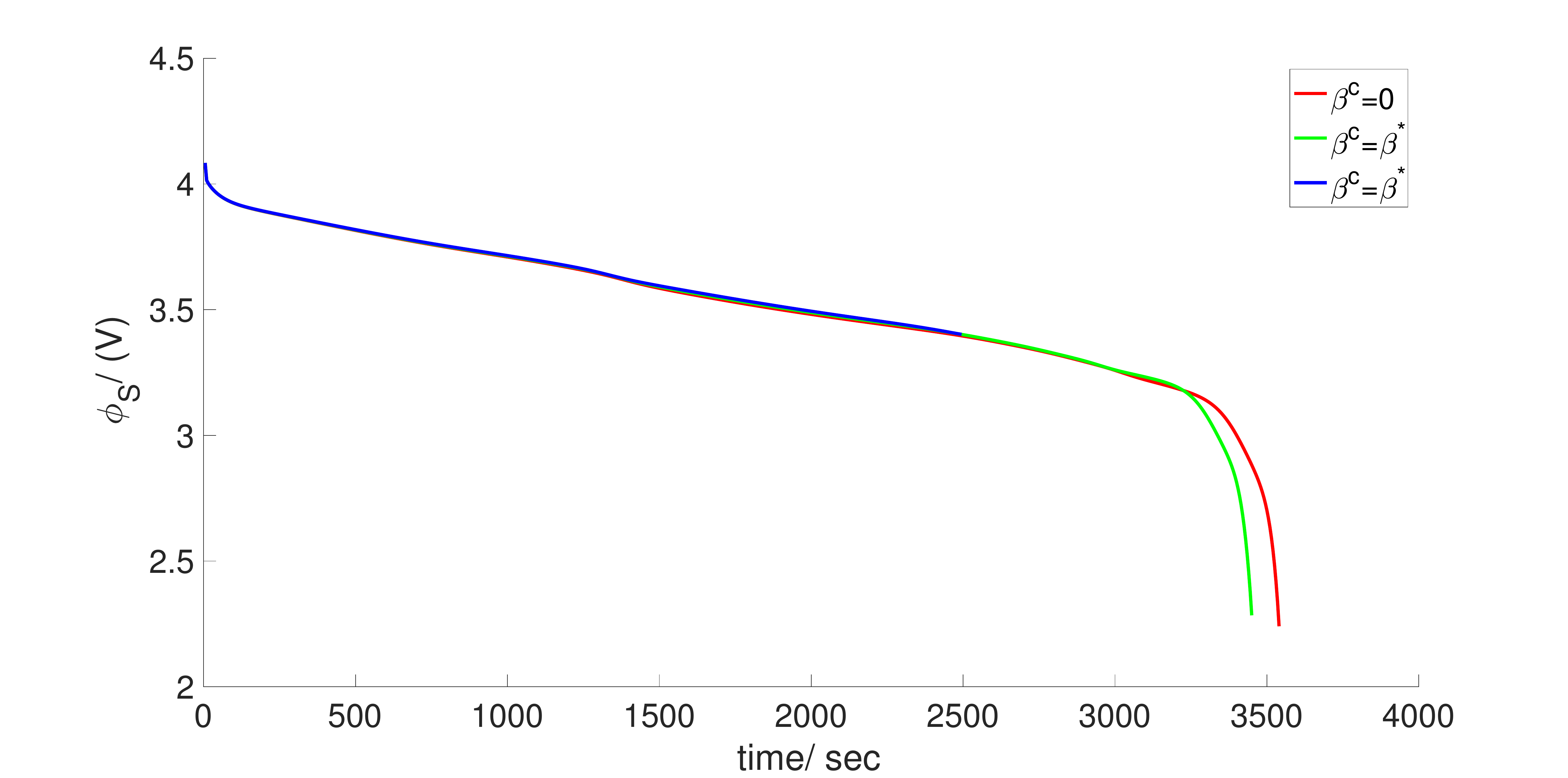}
\caption{\color{black} Electric potential, $\phi_\text{S}$ at Surface 2 decreases during discharge.}
\label{fig:phis_beta} 
\end{figure}

\subsection{Results for microstructures with different particle sizes}
We compute a discharge semi-cycle with the microstructures shown in Figure \ref{fig:ParticleModel}.  The boundary conditions correspond to Case 1. Note that, assuming uniform conditions initially, the time needed for fully discharging the cell could be calculated as 
\begin{equation}
\bar{t}=\frac{C_\text{Li}^0\epsilon_\text{s0}V_0}{S_\text{ext}\bj_\text{ext}}
\end{equation}
where $C_\text{Li}^0$ is the initial lithium concentration, $\epsilon_{\text{s}_0}$ is the initial volume fraction of active material, $V_0$ is the total volume of the electrode and $S_\text{ext}$ is the area of the boundary surface where the external current flux, $F\bj_\text{ext}=\bi_\text{ext}$, is applied. The external current was applied  such that all cases discharged fully over the same time. {\color{black} }

Figure \ref{fig:theta_R} shows the temperature increasing much faster for cases with larger particle size at the same porosity. This is the effect of smaller specific areas {\color{black} (see Figure \ref{fig:ParticleModel})}, and therefore higher reaction rates needed under the imposed currents with larger particles. The currents are different for different particle sizes, but even accounting for that, the reaction rates are higher for larger particles to maintain those currents with smaller specific areas.
\begin{figure}[hbtp]
\centering
\includegraphics[scale=0.2]{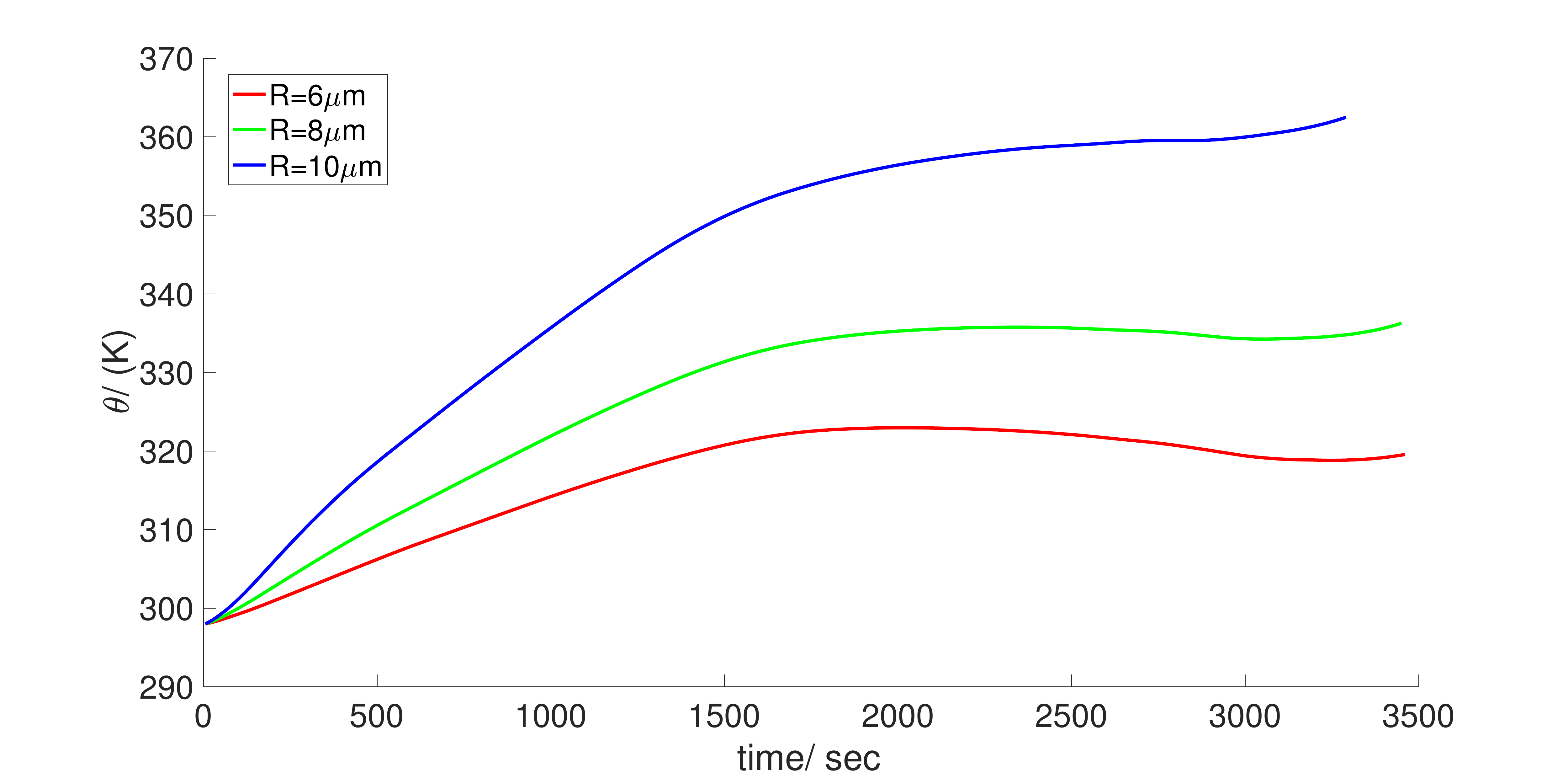}
\caption{\color{black}Temperature profile with three different particle sizes.}
\label{fig:theta_R} 
\end{figure}
Due to these effects involving specific areas and applied currents, with larger particle size the lithium decomposition rates are higher over the negative electrode and recombination rates  are higher over the positive electrode as shown in Figure \ref{fig:C_li_R}.
\begin{figure}[hbtp]
\centering
\includegraphics[scale=0.1]{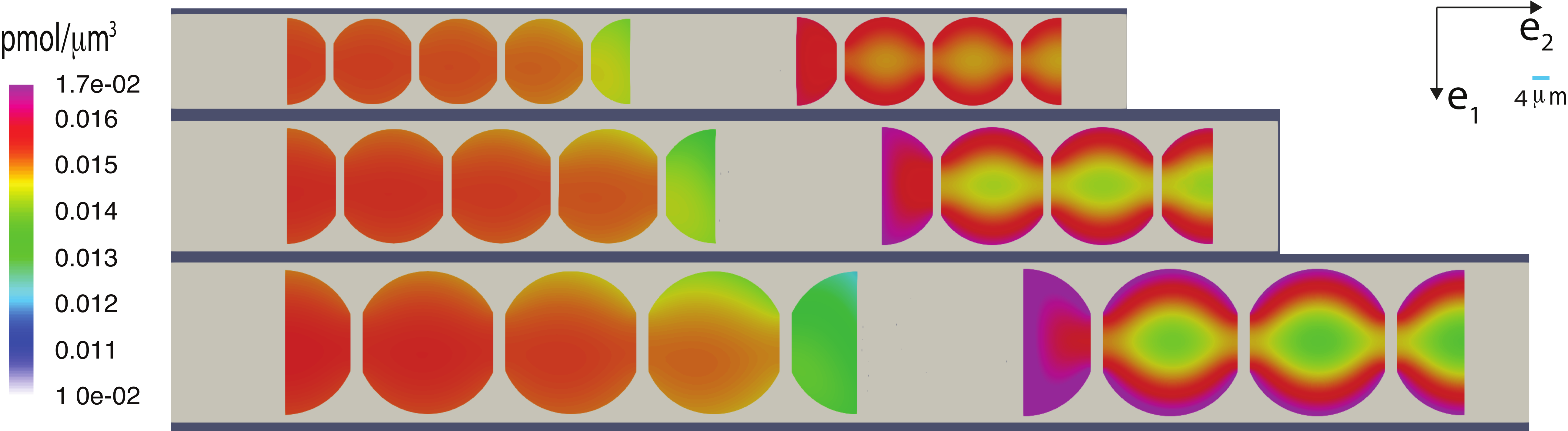}
\caption{\color{black}Lithium concentration in the deformed configuration with three different particle sizes: $6\;\mu$m, $8\;\mu$m, $10\;\mu$m from top to bottom.}
\label{fig:C_li_R} 
\end{figure}
Lithium ions are produced at the negative electrode and consumed at the positive electrode. Consequently we observe that the lithium ion concentration is distributed non-uniformly along the cell direction (as shown in Figure \ref{fig:C_li_plus_R}). This non-uniformity is exaggerated with larger particle sizes {\color{black} due to {\color{black}the thicker electrode, and }the higher rates of decomposition and recombination at the applied currents, as explained above.}
\begin{figure}[hbtp]
\centering
\includegraphics[scale=0.1]{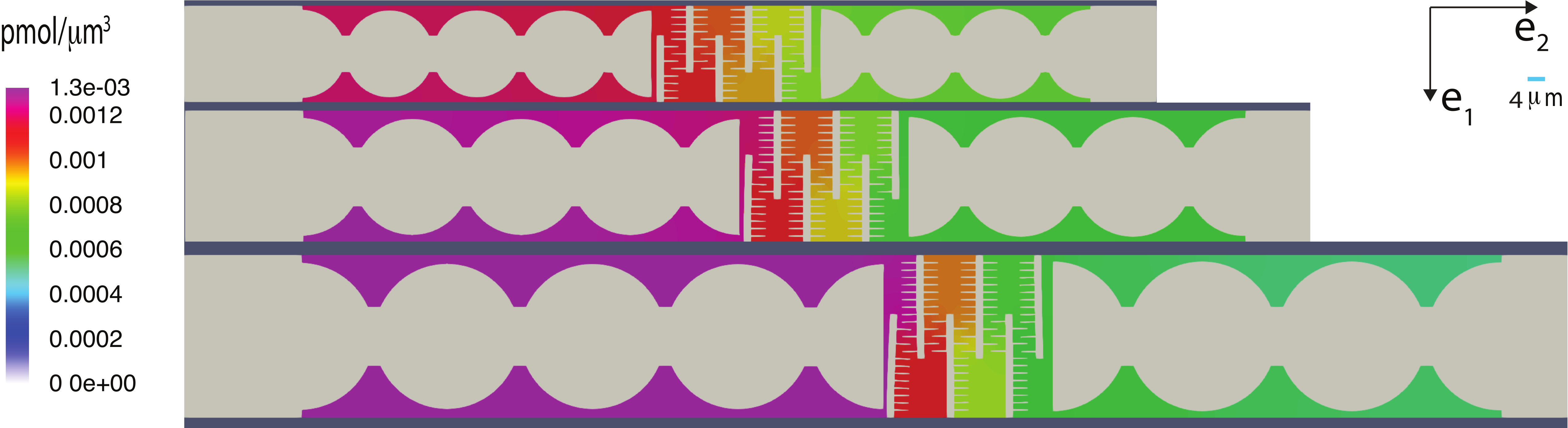}
\caption{\color{black}Lithium ion concentration in the deformed configuration with three different particle sizes: $6\;\mu$m, $8\;\mu$m, $10\;\mu$m from top to bottom.}
\label{fig:C_li_plus_R} 
\end{figure}
The higher rates for the cases with larger particle size also translate to the porosity changing slightly faster as shown in Figure \ref{fig:epl_R}, and the voltage also dropping slightly faster as shown in Figure \ref{fig:phis_R}.
\begin{figure}[hbtp]
\centering
\includegraphics[scale=0.2]{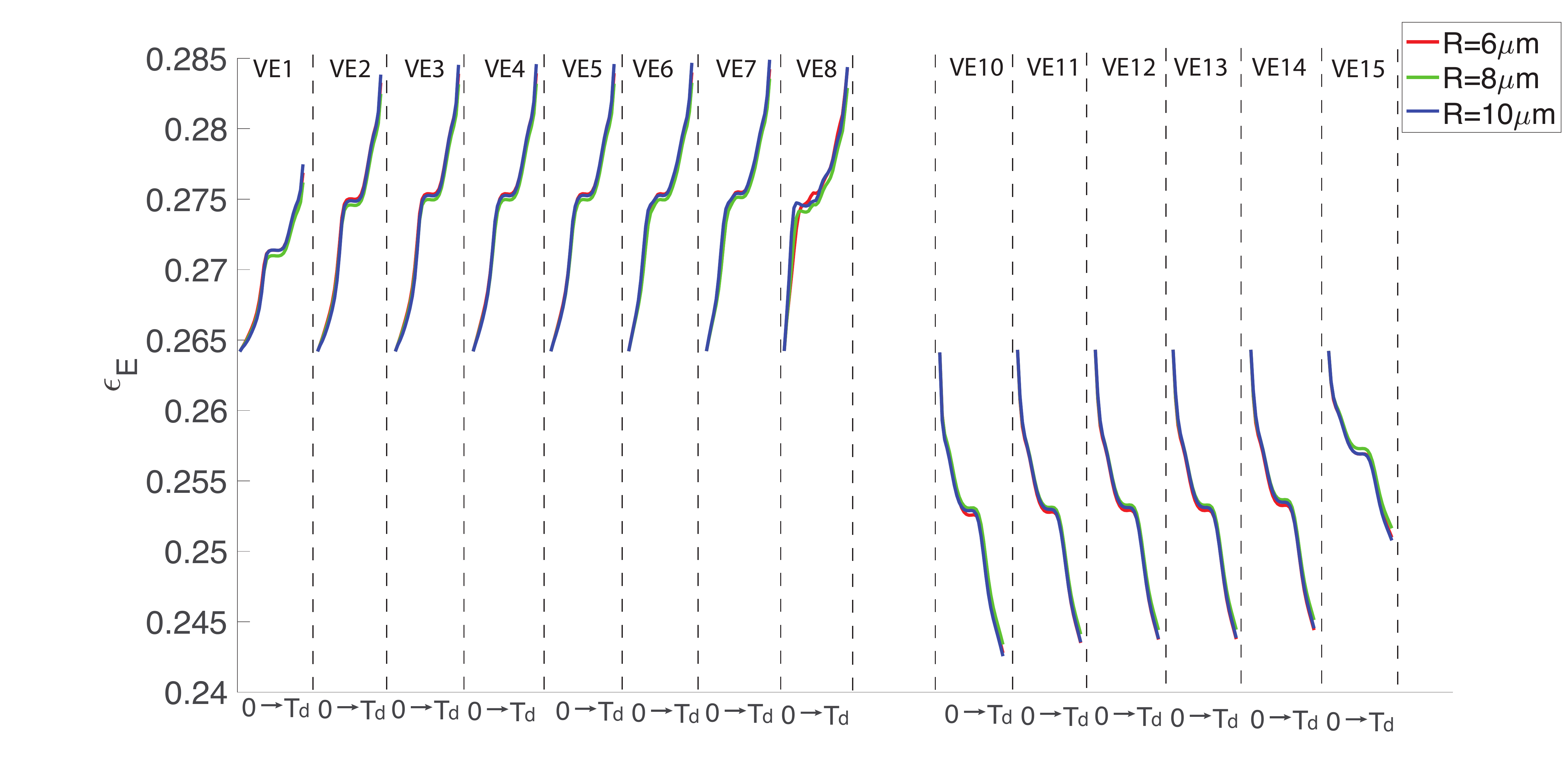}
\caption{\color{black}Porosity changes of each volume element in the electrode, for microstructures with three different particle sizes.}
\label{fig:epl_R} 
\end{figure}
\begin{figure}[hbtp]
\centering
\includegraphics[scale=0.2]{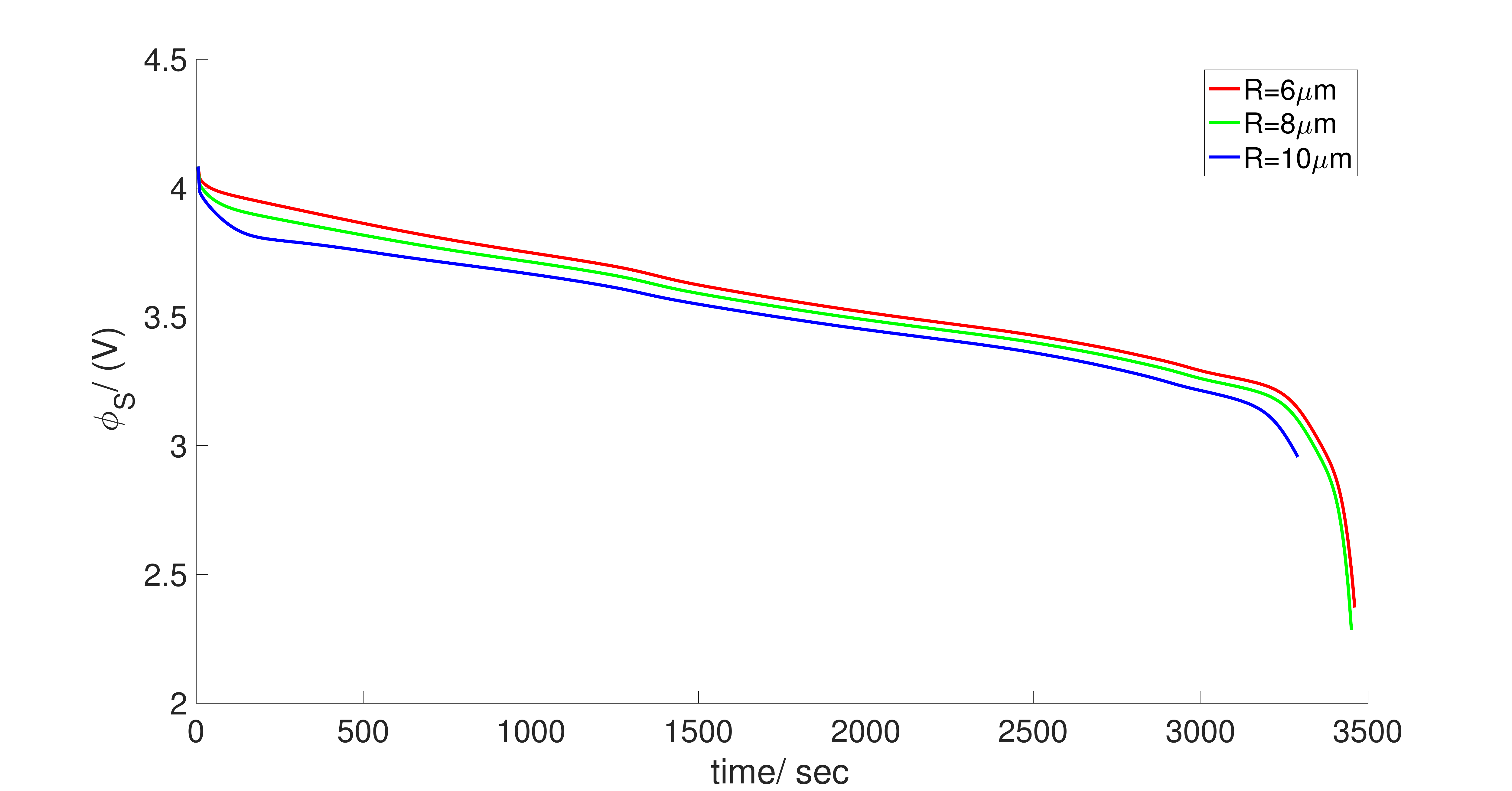}
\caption{\color{black}Electric potential, $\phi_\text{S}$ at Surface 2 decreases at a higher rate during discharge for microstructures with larger particle sizes. 
}
\label{fig:phis_R} 
\end{figure}

\section{Discussion and Conclusions}
\label{sec:Conclusions}
During battery charging (discharging), lithium intercalation (de-intercalation), thermal and elastic strains, drive the deformation of the active material, which in turn drives flows of the surrounding electrolyte. While the electrolyte flow velocity is too small (Figure \ref{fig:Pa_v}) to have a significant influence on lithium ion transport (Figures \ref{fig:Pa_cli} and \ref{fig:Pa_cli+}), the solid-fluid interaction affects the deformation of the solid phases significantly. The resultant evolution of porosity is strongly affected by far-field mechanical boundary conditions (Figures \ref{fig:epl}). For a given initial porosity, the different far-field boundary conditions have insignificant effects on the battery's performance during discharge provided the intercalation strain is small (see Figure \ref{fig:phis}).

The intercalation strains affect not only the microstructures of porous electrode, but also the microstructure of the separator through solid-fluid interaction (see Figure \ref{fig:C_li_plus_beta}). During battery operation, lithium ions undergo transport through the microstructure of the polymeric separator. As the intercalation strain function is scaled upward, there is more pronounced contraction and expansion of the porous microstructure. This affects the battery's overall performance, including its voltage response (see Figure \ref{fig:phis_beta}). Large intercalation strains may even cause pore closure, and failure of the battery.

Further insight comes from studies with different particle sizes. The evolution of porosities (Figure \ref{fig:epl_R}), and of the volume element-averaged lithium and lithium ion concentration (Figures \ref{fig:C_li_R} and \ref{fig:C_li_plus_R}, respectively) are strongly influenced by particle sizes if discharge rates are held fixed: Larger particles, having smaller specific areas for reaction, have higher reaction rates, causing higher gradients in porosity, lithium and lithium ion concentration. As expected, larger particle sizes translate to larger potential drops over the cycle. 

A comprehensive study of the trade-off among different microstructures,  including the consideration and evolution of porosity and tortuosity of these microstructures, also can be carried out by the framework presented here.  More promisingly, a data driven model based on large-scale direct numerical simulation data may be proposed, which could provide insights to battery design. The studies in this communication present an approach to generate large amounts of data via direct numerical simulations that can be used to parameterize complex response functions.
}
\section*{Acknowldegements}
This work was supported by Toyota Research Institute through the Accelerated Materials Design and Discovery program at University of Michigan.

\section*{Appendix}
\label{sec:Appendix}

{\color{black}
\subsection*{Demonstration of the formulation in three-dimensions}
In this section, we simulate a discharge process in three dimensions. While a detailed three-dimensional study of evolving particle-scale porosity, solid-fluid interaction, and their influence on battery performance is beyond the scope of this communication, this brief section serves to demonstrate the extensibility of our framework to three dimensions. It will be followed by a separate study in three dimensions. 

The configuration studied appears in Figure \ref{fig:3D_domain}. The boundary conditions correspond to Case 1, for which the displacement was fixed on all surfaces. Symmetry has been used to reduce the computational domain to one quarter of the physical one. Pending a large scale numerical computation, with its associated tuning of mesh densities, solver settings and computational resources, we have used a simpler, less resolved representation of the separator. This delivers a solution that demonstrates the three-dimensional capability, if not a high-fidelity simulation.
\begin{figure}[hbtp]
\centering
\includegraphics[scale=0.15]{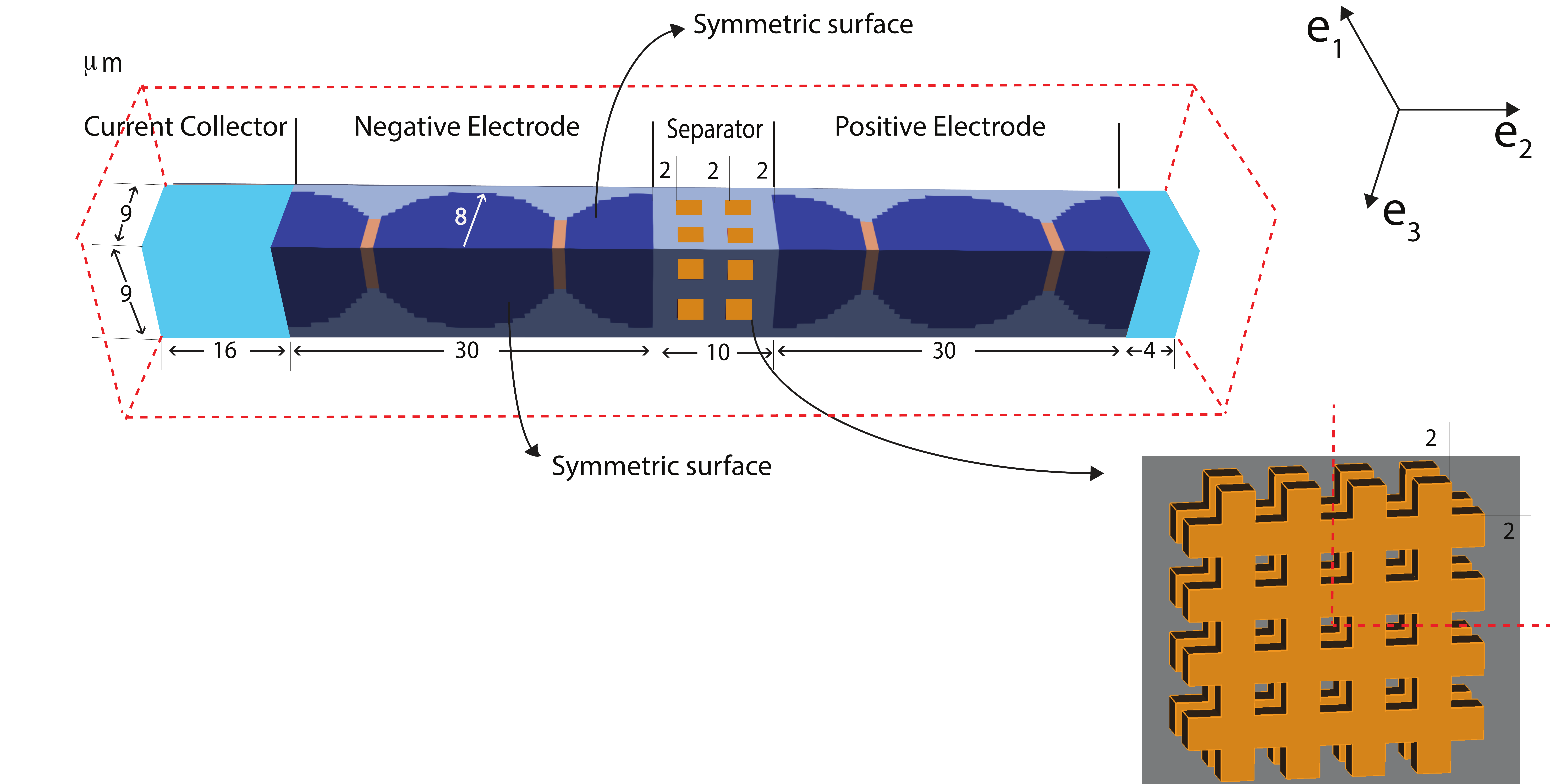}
\caption{{\color{black}Schematic of one quarter of a three dimensional configuration of active particles in the electrodes, with a simplified separator.}}
\label{fig:3D_domain} 
\end{figure} 

Figure \ref{fig:V_stress} shows the distributions of the velocity of the electrolyte, and von Mises stress, plotted in the deformed configuration at SOC = 0.97. This represents an early stage during discharge. The velocity magnitude is higher than that shown in the two-dimensional computation of Figure \ref{fig:Pa_v}, which however, corresponds to SOC = 0.5. The von Mises stress distribution is in better agreement with the corresponding two-dimensional result in Figure \ref{fig:Pa_vonMises}. The distribution of lithium is nonuniform in the radial direction within the active particle, and both lithium and lithium ion distributions are nonuniform along the cell as shown in Figure \ref{fig:3D_C}. The lithium and lithium ion concentrations are closer to the two-dimensional results in Figures \ref{fig:Pa_cli} and \ref{fig:Pa_cli+}. However, a detailed study will appear in a future communication.
\begin{figure}[hbtp]
\centering
\includegraphics[scale=0.15]{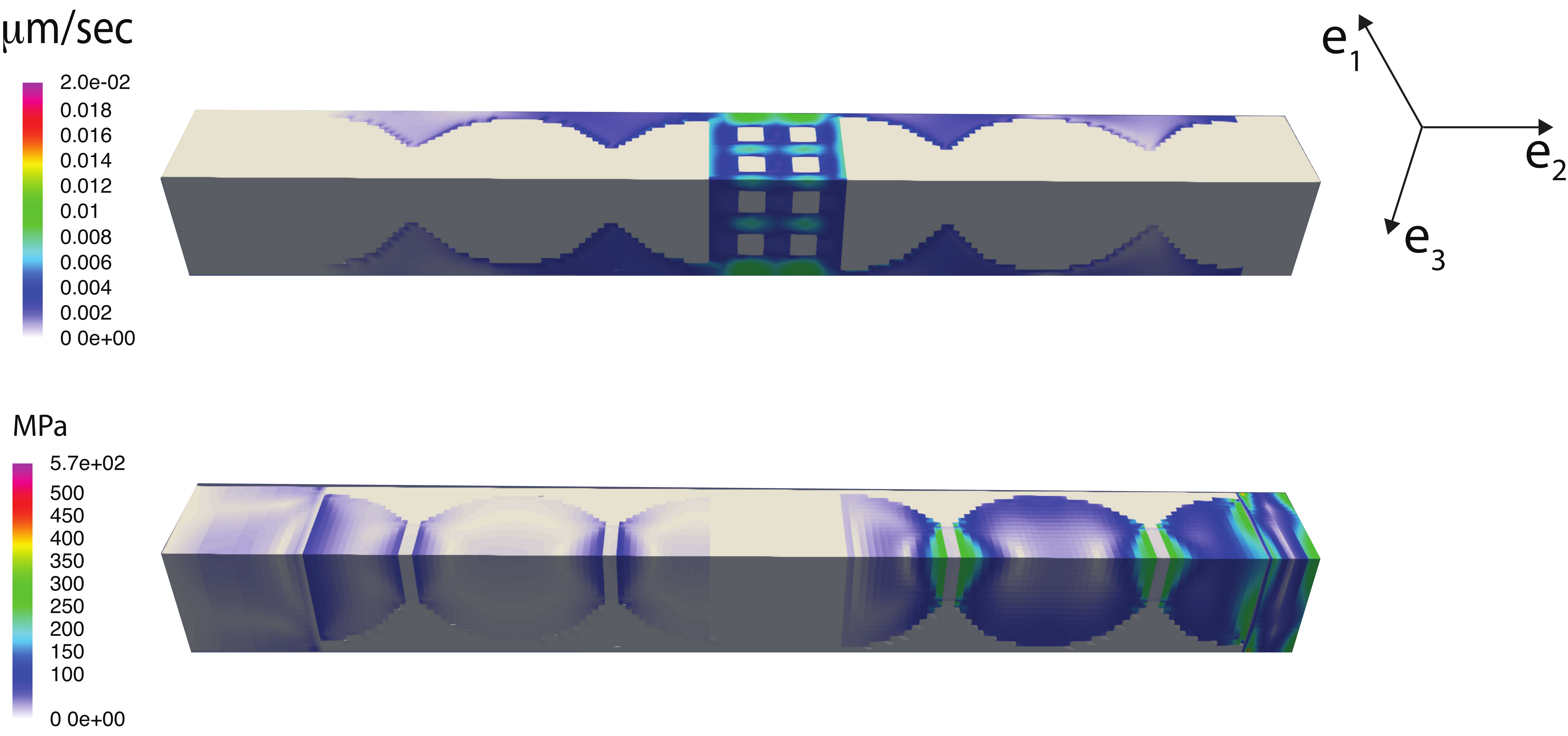}
\caption{{\color{black}Velocity of electrolyte(upper plot) and von Mises stress(lower plot) profile at the deformed configuration at SOC=0.97.   }}
\label{fig:V_stress} 
\end{figure}

\begin{figure}[hbtp]
\centering
\includegraphics[scale=0.15]{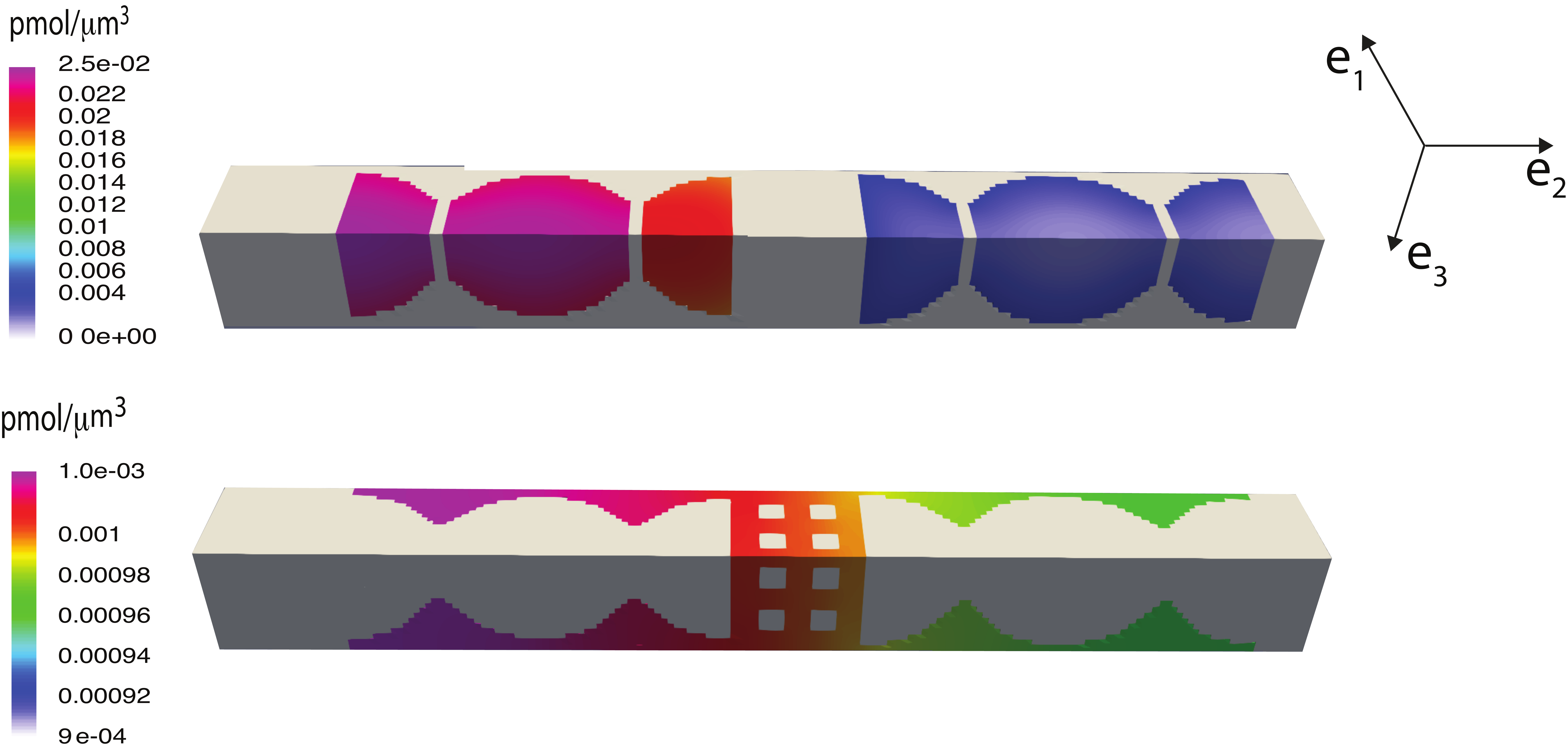}
\caption{{\color{black}Lithium (upper plot) and lithium ion (lower plot) distribution at the deformed configuration at SOC=0.97.   }}
\label{fig:3D_C} 
\end{figure} 
}
\subsubsection*{Intercalation strain function and diffusion coefficient}
The swelling function $\beta^\text{c}$ is obtained from\cite{ZWJES2017}:
\begin{align}
{\beta}^\text{c}(\eta) &=1.496\eta^3-1.739\eta^2+1.020\eta-0.033\exp(2.972\eta)-0.046\tanh(\frac{\eta-0.1}{0.1})\nonumber\\
              &-0.004\tanh(\frac{\eta-0.3}{0.1})+0.021\tanh(\frac{\eta-0.65}{0.1}).
\label{eq:beta_s}
\end{align}
where $\eta=C_\text{Li}/C_\text{Li}^\text{max}$.

The diffusion coefficient, $D_+$ is given by\cite{White2010JES}
\begin{align}
\log(D_+)=-4.43-\frac{54}{\theta-5\times 10^3C_{\text{Li}^+}-229}-2.2\times 10^2C_{\text{Li}^+}
\label{eq:D_plus}
\end{align}

\end{document}